\documentclass[10pt,aps,prx,showpacssuperscriptaddress,preprint,citeautoscript]{revtex4-1}

\usepackage{amssymb, amsmath,amsfonts}    

\usepackage{color}      
\usepackage{graphicx}	
\usepackage{subfigure}   
\usepackage{float}
\usepackage{verbatim} 
\usepackage{hyperref}
\usepackage[normalem]{ulem}

\DeclareGraphicsExtensions{.eps,.png,.pdf,.jpg}

\def\eeq{\relax}
\def\beq#1#2\eeq{\begin{equation}\label{#1}#2\end{equation}}
\def\bal#1#2\eal{\begin{align}\label{#1}#2\end{align}}
\def\bse#1#2\ese{\begin{subequations}\label{#1}#2\end{subequations}}
\def\ba{\begin{aligned}}   \def\ea{\end{aligned}}

\def\k0{{\bf k}_0}

\def\dd{\operatorname{d}} 
\def\ii{\operatorname{i}}

\def\sgn{\operatorname{sgn}} 

 \newcommand{\ann}[1]{{\textcolor{black}{#1}}} 
  
\begin{document}

\title{Total acoustic transmission between fluids using a solid material with emphasis on the air-water interface}

\author{Hesam Bakhtiary Yekta$^{1}$ and Andrew N. Norris }
\address{ Department of Mechanical and Aerospace Engineering, Rutgers University, Piscataway, NJ 08854, USA}
\date{\today}



\begin{abstract}

Total acoustic transmission between water and air is modeled using a purely solid interface comprising two elastic plates separated by periodically spaced  ribs.  The frequency of full transmission depends only on, and  is inversely proportional to,  the areal density of the plate facing the air.    Total transmission also requires a specific dependence of the  rib spacing  on the bending stiffness of the two plates.  These  relations are the  result of an explicit analytical solution for the transmitted and reflected acoustic waves combined with asymptotic approximations based on the small parameter defined by the air-to-water impedance ratio.  Surprisingly, the total transmission effect is almost independent of the angle of incidence, even though the transmission conditions are predicated on normal incidence. Parametric studies are performed to examine the effect on the frequency bandwidth and Q-factor of the acoustic transmissivity.  A lower bound for the Q-factor of  $30.6$ is simply related to the water-air impedance ratio. 

\end{abstract}

\maketitle

\section{Introduction} \label{sec1}   


In a 1942  US patent, Hansell \cite{Hansell} first described how impedance matching between    acoustic fluids can be achieved by  a quarter wavelength  intermediate layer  with impedance equal to the harmonic mean of the two media.  The layer results in  zero reflection and total energy transmission.  Hansell also suggested a two layer solution, also quarter wavelength, with each impedance the harmonic mean of its neighbors; the goal of introducing two or more layers is to increase the bandwidth of the single layer device.   Hansell's ideas saw little immediate application in acoustics; rather, the focus of impedance matching in the mid-20th century was on microwave transformers, with a comprehensive body of literature, e.g. \cite{Southworth1950}. 
A later  acoustics related  application of impedance layers emerged for piezoelectric transducers in contact with air or water. In this case, the finite thickness of the piezoelectric element should be considered leading to, for instance, a single layer impedance $Z_\text{air}^{2/3}Z_\text{piezo}^{1/3}$ \cite{Desilets1978}.  Here we are concerned with energy transmission between semi-infinite acoustic media so that  Hansell's single layer impedance is appropriate \cite{Hansell}.   The difficulty lies in finding the specific material with the desired intermediate impedance.  In particular, no naturally occurring material has the required impedance for the air-water interface. 

Recent  interest in impedance matching between air and water has been prompted by the innovative   approach of  Bok et al.\  \cite{Bok2018} that employs an air layer as a spring in series with a membrane mass.   This and most other  proposed methods require air-water interfaces  either through membranes  \cite{Bok2018},  bare bubbles \cite{Bretagne2011,Bolghasi2017,Cai2019,Lee2020}, bubbles within a membrane \cite{Gong2023}, hydrophobic materials \cite{Huang2021a}, lotus acoustic metasurface \cite{Huang2021Lotus}, air channels \cite{Liu2023}, metal inclusions in air and in water \cite{zhou2023gradient}.    Another proposed method uses airborne radar to measure the water surface acoustic displacement \cite{Tonolini2018}.     \ann{Other  approaches include}  \cite{Zhou2023,Zhang2024} which uses a layer of 3D printed epoxy designed via topology optimization to have impedance equal to the harmonic mean of water and air, 
\ann{and transmission at a "fish scale" resonance \cite{Esposito2024}. }

We present a simple acoustic impedance matching mechanism for the water-air interface. The  matching layer, or transformer, is made from a solid material, e.g.\ aluminum.  No fluid layers, either water or air,  membranes or other mechanisms are required.   Unlike the other proposed transformers including  the solid ones \cite{Zhou2023,Zhang2024,Zhou2023broadband} the present approach is analytically explicit, with closed form expressions relating the performance characteristics, e.g. transmission frequency, to the material properties.  This allows us to find  remarkably simple relations  by taking advantage of the  asymptotically small parameter defined by the ratio of the air and water acoustic impedances.  

By way of introduction to the problem we begin in Section \ref{sec2} with a simple but fundamental transformer  model motivated by the ideas of Bok et al.\  \cite{Bok2018} and others  \cite{Huang2021a,zhou2023gradient}.  In  the process  we show why  models with air-water interfaces are   very  difficult to realize in practice.   The  flex-layer impedance transformer  is presented in \ann{Section \ref{sec3}} along with a summary of the main results.   Mathematical details are given in Section \ref{sec4}, and the conditions for total transmission are derived in Section \ref{sec5}.  The efficacy of the proposed model is demonstrated in Section \ref{sec6} along with discussion of factors  such as the effect of  entrained air between the plates, and concluding  with a  comparison  with the fundamental  one-dimensional model presented next.  A summary and conclusions are given in 
 Section \ref{sec7}.
 
\section{ A simple water-to-air impedance matching using water and air layers} \label{sec2} 

  \begin{figure}
  \centering
    \includegraphics[width=0.66\textwidth]{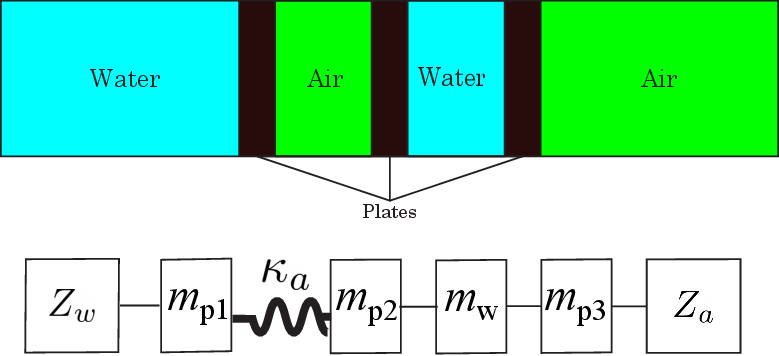}
  \caption{A spring-mass resonator separating semi-infinite water on the left and air on the right. The air layer acts as the spring $\kappa_a$, and the water layer  is a mass $m_\text{w}$,.  The dividing thin plates  add masses $m_\text{p1}$,$m_\text{p2}$, and $m_\text{p3}$, to the system. }
	\label{fig1}
   \end{figure}

The acoustic  properties of air and water are characterized by densities, $\rho_a$, $\rho_w$, and 
bulk moduli  $K_a$, $K_w$,  
with derived quantities, sound  speed $c = \sqrt{{K}/\rho}$ and impedance $Z = \sqrt{K \rho}$. 
 Figure \ref{fig1} depicts the   model:   semi-infinite bodies of  water and  air surround a transformer comprising thin layers of air and water,
with thin plates separating them as
 \textcolor{blue}{w}\,\textcolor{red}{{\vrule width 1pt}}\,a\,\textcolor{red}{{\vrule width 1pt}}\,\textcolor{blue}{w}\,\textcolor{red}{{\vrule width 1pt}}\,a.  
The central \textcolor{red}{{\vrule width 1pt}}\,a\,\textcolor{red}{{\vrule width 1pt}}\,\textcolor{blue}{w}\,\textcolor{red}{{\vrule width 1pt}} transformer has mass elements  $m_\text{p1}$, $m_\text{p2}$,  $m_\text{p3}$, and $m_\text{w}$ for the plates and the water layer, and  spring element $\kappa_a$ for the air layer.  
\ann{Assuming time dependence $e^{-\ii \omega t}$ with $\omega = 2\pi f$, }
the effective impedance of the transformer in series with the semi-infinite air can be calculated using  low frequency lumped parameters models \cite{blackstock2001fundamentals} as  
\beq{-1} Z = -\ii \omega m_\text{p1} +   \big\{  (Z_a - \ii \omega m)^{-1} - \ii \omega \kappa_a^{-1} \big\}^{-1} 
\eeq
where $m = m_\text{w} +  m_\text{p2}+ m_\text{p3}$. The  condition for 
 impedance matching to water,  $Z = Z_w + 0\ii $,  becomes, taking $ m_\text{p1}=0$ for simplicity since it acts only as a phase term in water, 
\beq{1}
 \kappa_a m = Z_w Z_a, 
\quad \text{and} \quad 
 \omega \approx \omega_0 \equiv  \frac{\sqrt{Z_w Z_a }}m. 
\eeq
Condition \eqref{1}$_1$ says the transformer impedance $\sqrt{\kappa_a m}$ is the geometric mean of $Z_w$ and $ Z_a$ in agreement with   Hansell \cite{Hansell}.  Condition \eqref{1}$_2$  follows from the exact solution 
$\omega = \omega_0 \, (1-\epsilon)^{1/2}$   based on  asymptotics for the  small parameter 
\beq{3+0}
\epsilon \equiv \frac{Z_a}{Z_w}  = 0.267\times 10^{-3}, 
\eeq
and  shows that full transmission is at the resonance frequency  $\omega_0
= \sqrt{\kappa_a /m}$.     

  \begin{figure}
  \centering
    \includegraphics[width=0.9\textwidth]{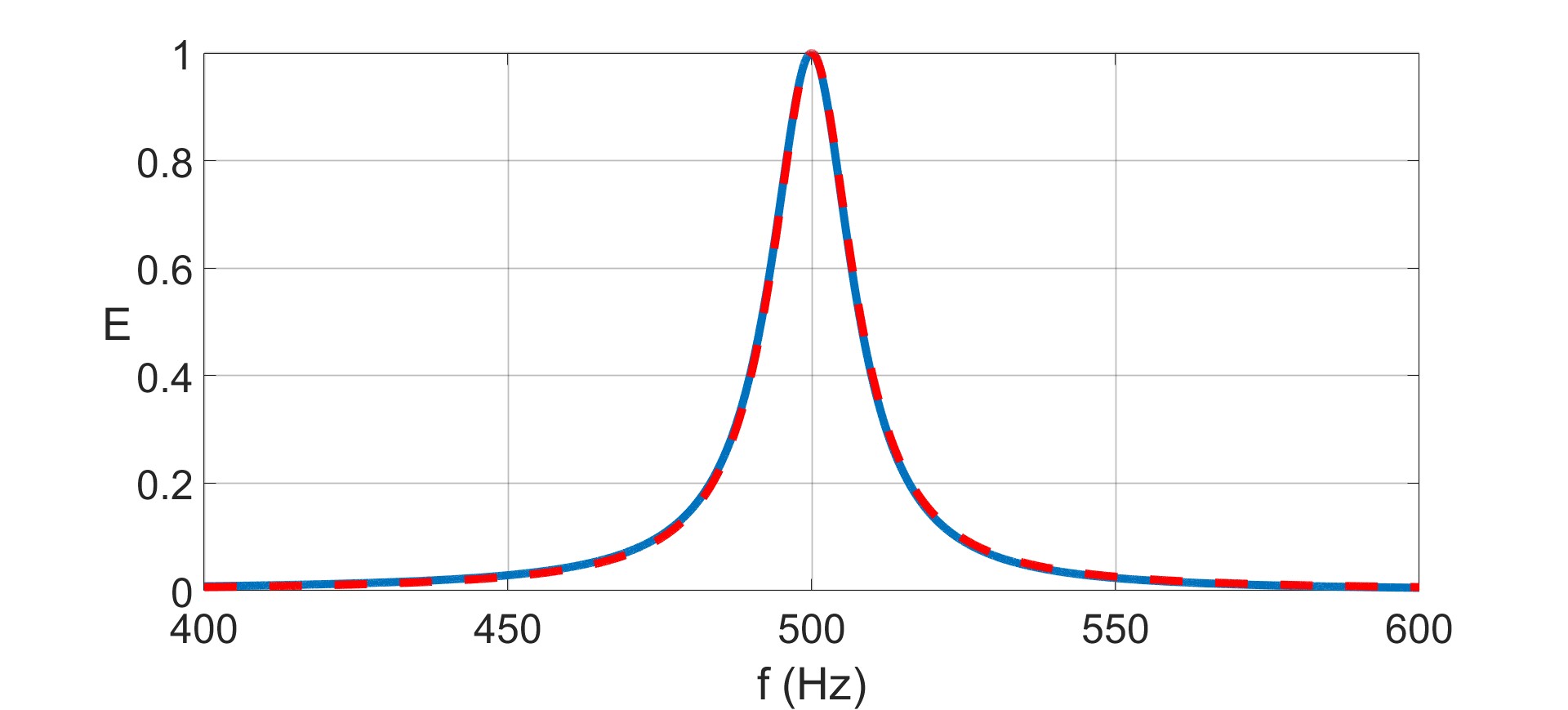}
  \caption{ Transmitted energy for unit incident energy  from the water side, $f_0=500$ Hz. Solid curve is a full wave (exact) simulation and the dashed curve is the \ann{asymptotic} approximation E\, $ \approx 1/\Big(1 + \frac 1{\epsilon} \big( \frac{f}{f_0} -1  \big)^2   \Big)$.  \ann{The latter follows from the lumped-parameter impedance of Eq.\ \eqref{-1} as  E\, $=1 - |R|^2$ for reflection coefficient 
  $R = \frac{Z - Z_w}{Z + Z_w}$.  }}
  \label{fig2}
   \end{figure}
Figure \ref{fig2} shows a full wave simulation using transfer matrices for three 1 mm thick Al plates.
Dissipation in the air layer from viscous and thermal diffusivity is included 
\cite{Lighthill80} with  no evident influence. 
Two curves are  plotted, solid and dotted: the latter is an asymptotic approximation for the transmitted energy in air based on the lumped parameter model,  E\, $ \approx 1/\Big(1 + \frac 1{\epsilon} \big( \frac{f}{f_0} -1  \big)^2   \Big)$.   Its remarkable accuracy indicates how there are no adjustable quantities in the response, even though we have built  free parameters into the design.  
This simple formula for E says that the $Q$-factor of the transmission resonance, which arises from radiation damping, not from energy dissipation, has a characteristic value of $Q \approx  \frac 1 {2\sqrt{\epsilon}} = 30.6$.   We will return to this value later in \S   \ref{sec6b}. 

Introduce effective thicknesses $d_a$ and $d_w$ for air and water according to 
\beq{2}
\kappa_a = \frac{\rho_a c_a^2}{d_a}, 
\qquad 
m = \rho_w d_w ,
\eeq
then  \eqref{1} translates to  the  equivalent conditions 
\beq{3}
\frac{\omega_0 d_a}{c_a} = \sqrt{\epsilon}, 
\qquad
 \frac{ d_w}{c_w}  = \frac{ d_a}{c_a} . 
\eeq
The fact that $\sqrt{\epsilon} = 0.016$ justifies the subwavelength approximations. 
While the thickness of the effective water layer is adjustable, based on the choice of the plates  which provide mass, the thickness of the air layer  is severely limited  by the fact that it is the only stiffness operating, and Eq.\ \eqref{3}$_1$ constrains it as $d_a = 0.87/f_0$ m at $f_0$ Hz transmission frequency. 
This requires plate separation $d_a$ on the order of 1 mm at 870 Hz and less for higher $f_0$ which seems hard to imagine over large areas in  water.  Placing  spacers between the air plates would keep the plates separated, but  adds  stiffness in parallel which  requires $d_a$ to be further decreased.  This is not a feasible solution since any parallel stiffness would overwhelm the air stiffness. While it is instructive, the simple model of Fig.\ \ref{fig1} is impractical. 

An alternative approach is discussed next  which  replaces the fluid  \textcolor{red}{{\vrule width 1pt}}\,a\,\textcolor{red}{{\vrule width 1pt}}\,\textcolor{blue}{w}\,\textcolor{red}{{\vrule width 1pt}} transformer with a purely solid   \emph{flex layer}.

\section{Problem setup and summary of the main results} \label{sec3}   

The impedance transformer is motivated by the fact that     a pair of parallel  elastic plates in water and separated by periodically spaced ribs, a "flex-layer", acts at low frequency as an equivalent stiffness  \cite{BakhtiaryYekta2024}.  This contrasts with the usual low frequency approximation of an elastic solid as a mass, leading to the well known mass-law transmission loss \cite[Sec.\ 6.7]{KinslerFrey}.
The effective stiffness of the flex-layer offers an alternative to the air-layer stiffness of the simple model of  Fig.\ \ref{fig1}, while the mass of the plates provides the necessary mass for the transmission resonance. The flex-layer considered  here may have two distinct plates, but as we will see, only the one facing the air contributes to the effective mass of the resonator.  

\subsection{The flex-layer impedance   transformer } 

\begin{figure}
    \centering
    \includegraphics[width=0.7\textwidth]{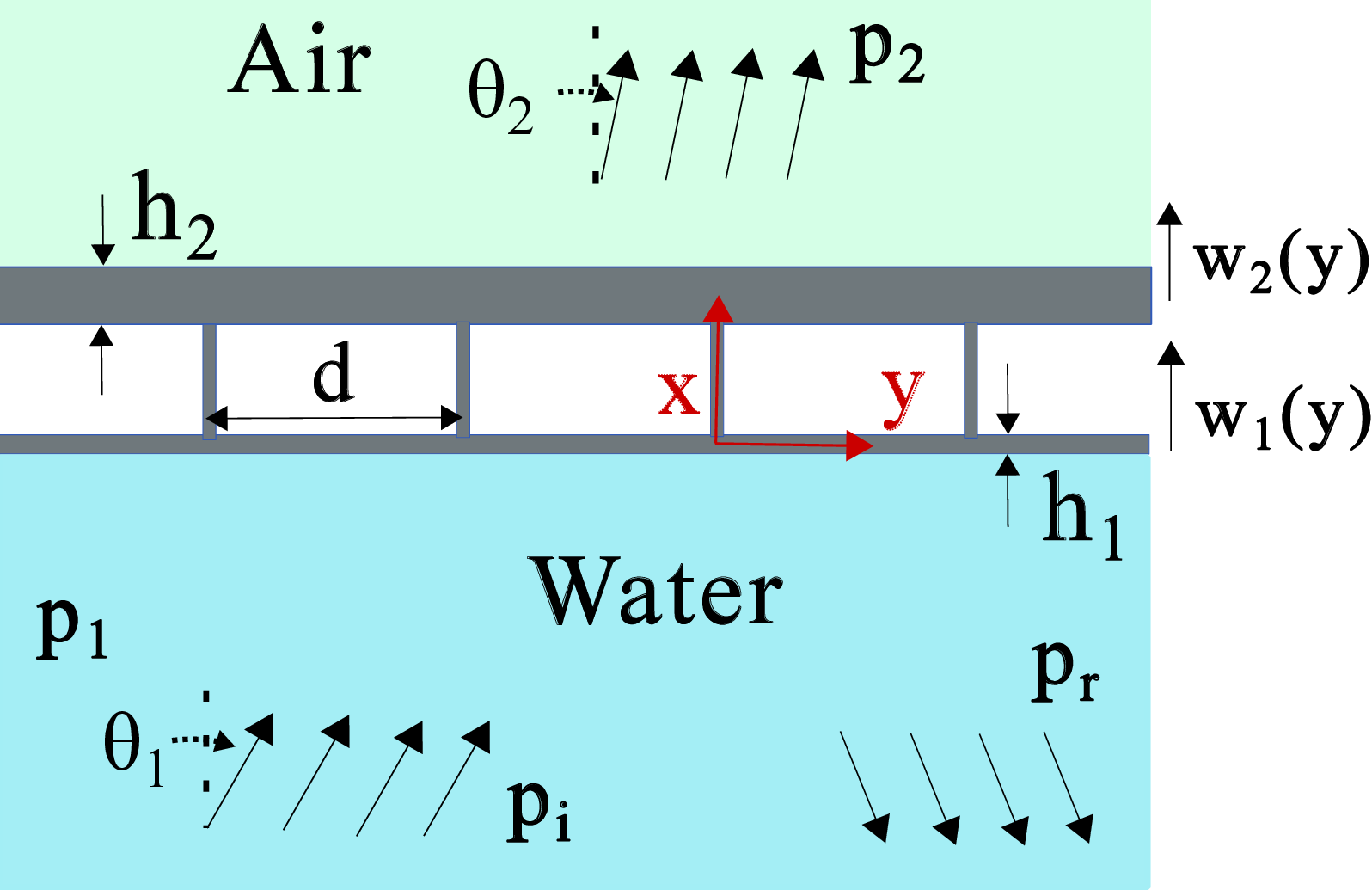}
    \caption{A plane wave is incident from the water side of the asymmetric panel. The plates are separated by ribs set a dsitance $d$ apart with the intermediate space assumed to be vacuum.  The effect of entrained air is small and is discussed in Section \ref{sec6d}.}
    \label{Flex_thin_thick}
\end{figure}

The transformer layer is depicted in Fig.\ (\ref{Flex_thin_thick}). 
The semi-infinite water and air regions occupy $x<0$ and $x>0$, respectively  (the finite gap between the plates is compressed into a single point for simplicity). 
We label water and air quantities with subscripts $1$ and $2$, respectively, so that the wavenumbers are $k_j =\omega /c_j$ where $c_j$ are the sound speeds, and for later $\rho_j$ are the densities.
We consider plane wave incidence in water at angle 
$\theta_1 $ from the normal, with $y-$wavenumber $ k_0 = k_1 \sin\theta_1$.   
The fundamental transmitted wave in air is at angle $\theta_2$  which follows from Snell's law: $k_2 \sin\theta_2 =k_1  \sin\theta_1$, and hence $\theta_2 \le \theta_1$.  

 The impedance matching system comprises two parallel plates  of thickness $h_1$ and $h_2$
 separated by  periodically spaced ribs, see Fig.\ \ref{Flex_thin_thick}.  
 The plate parameters are mass  per unit area   $m_j = \rho_{sj} h_j $ and  bending stiffness 
  $D_j = E_jI_j/(1-\nu_j^2)$ where $I_j = \frac{h_j^3}{12}$, $j=1,2$. 

\subsection{Principal results}

Here we summarize the main results that define the conditions required for total transmission under normal incidence $(\theta_j=0, j=1,2)$.  The first condition necessary  for full transmission relates the transmission frequency $\omega_0$ to the mass densities $m_1$, $m_2$, and the acoustic impedances $Z_j =\rho_j c_j$, $j=1,2$:
\beq{-21}
Z_1 + \frac{m_1^2\omega_0^2}{Z_1} =  Z_2 + \frac{m_2^2\omega_0^2}{Z_2} , 
\eeq
see \S \ref{sec5a} and Appendix \ref{appb}.
For the water-air interface $Z_1=Z_w$, $Z_2=Z_a$, this reduces to 
 \beq{44-0}
 \omega_0 \approx \frac{\sqrt{Z_wZ_a}}{m_2} 
 \eeq
 which is precise if $m_1=m_2$ and is otherwise less than $0.1$\% in error.  It is interesting to compare this with the  condition  \eqref{1}$_2$ for the simple model of Fig.\ \ref{fig1}, which also depends upon the mass facing the air half-space.  
 
 The second condition  involves solving for the zero of a nonlinear function, although the solution can be  approximated (see Section \ref{sec6}) for the air-water interface by a   relation similar to  \eqref{1}$_1$ for  the simple transformer model
  \beq{059-}
  \kappa \,m_2 \approx Z_a Z_w
  \quad \text{where} \quad 
\kappa =  \frac{720}{  d^4}  \Big( \frac 1{D_1} +   \frac 1{D_2}  \Big)^{-1}  .
 \eeq
 In practice,  \eqref{059-} provides the condition for determining the rib spacing $d$  that yields  full transmission at $\omega=\omega_0$ of \eqref{44-0}.  In summary, 
 the flex-layer acts as a spring-mass transformer with stiffness $\kappa$ that depends on the stiffness of both plates and mass $m=m_2$ that depends only on the mass of the plate on the air side.

\section{Scattering analysis for an asymmetric  panel}  \label{sec4}   

On the water side, plate 1 $x<0$, we consider the incident acoustic pressure $p_i$ along with its rigidly reflected pressure, which together  give zero normal velocity  on the plate.  The plate normal velocity, $v_1(y) = v_x(-0,y)$ is  therefore related to the additional pressure $p_1$  by the momentum equilibrium equation in the $x-$direction: $\ii \omega \rho_1 v_1(y) = \frac{\partial p_1}{\partial x}(0,y)$.  On the air side the total acoustic pressure $p = p_2$  radiates away from the plate in the positive $x-$direction, and the plate normal velocity, $v_2(y) = v_x(+0,y)$, is given by  $\ii \omega \rho_2 v_2(y) = \frac{\partial p_2}{\partial x}(0,y)$.
Introducing the $y-$transforms, 
\beq{7=2}
\hat V_j(\xi) = \int_{-\infty}^\infty v_j(y)  e^{-\ii \xi y} \dd y, 
\qquad
v_j(y) = \frac 1{2\pi}  \int_{-\infty}^\infty \hat V_j(\xi)  e^{\ii \xi y} \dd \xi , \quad j=1, 2 ,
\eeq   
it follows that the additional scattered pressure in the water $(j=1)$ and the total pressure in the air  $(j=2)$ are related to the normal velocities by 
\beq{7=3}
p_j(x,y) =  \frac {\sgn x}{2\pi}  \int_{-\infty}^\infty \hat Z_{fj}(\xi) \hat V_j(\xi)  
e^{\ii ( \sqrt{k_j^2-\xi^2}\, |x| + \xi y )}\, \dd \xi, \quad j=1, 2 
\eeq
with fluid  impedances   
\beq{7=4}
\hat Z_{fj}(\xi)= \frac{\rho_j \omega}{ \sqrt{k_j^2-\xi^2}  } , \quad j=1, 2 .
\eeq
The square roots in eqs.\  \eqref{7=3} and \eqref{7=4}  are either positive real or positive imaginary. 
In summary, the total pressure in water $(x<0)$ and air $(x>0)$
is
\beq{-44}
p(x,y) = \begin{cases} p_1(x,y) +
p_0 e^{ \ii k_1(x \cos\theta_1 +y \sin \theta_1)} + p_0 e^{\ii k_1(-x \cos\theta_1 +y \sin \theta_1)}
    , & x<0, 
    \\
    p_2(x,y), & x>0 ,
\end{cases}
\eeq
with $p_1$ and $p_2$ given by eq.\ \eqref{7=3}.

\subsection{Solution} 
The displacements of plates 1 and 2 in the $x-$direction, $w_j(y) = (-\ii \omega)^{-1}v_j(y)$, satisfy
\bse{4=78}
\bal{plate1_dis}
  \mathcal{L}_1 w_1(y) &=
 2p_0  e^{\ii k_0 y }  + p_1(0,y) 
  - \Big[   Z_{0+} (v_2+v_1)(y) - Z_{0-} (v_2-v_1) (y)  \Big]
 \sum_{l=-\infty}^\infty \delta(y- ld),
 \\
  \mathcal{L}_2 w_2(y) &=
 - p_2(0,y) 
  - \Big[   Z_{0+} (v_2+v_1)(y) + Z_{0-} (v_2-v_1) (y)  \Big]
 \sum_{l=-\infty}^\infty \delta(y-ld), 
 \label{plate2_dis} 
\eal
\ese
where $ \mathcal{L}_j w(y) =  D_j w''''(y) - m_j \omega^2 w(y) $, $j=1,2$. 
The stiffness impedance $Z_{0-}$  defines the force between the plates that depends on their relative separation, while $Z_{0+}$ is a mass impedance that depends on the motion of the rib center of mass.  \textcolor{black}{For simplicity in the equations, it is assumed that the ribs make point contacts with the plates as their thicknesses are negligible in comparison with the plate length.} Two models for $Z_{0\pm}$ are given in Appendix \ref{appa}. 
Substituting the  Poisson summation identity \cite{evseev1973sound,Lin1977}
\beq{7=11}
\sum_{l=-\infty}^\infty    \delta(y-l d)
 = \frac 1d   \sum_{m=-\infty}^\infty e^{-\ii 2\pi m \frac yd}
\eeq
and taking the $\xi$ transform of \eqref{plate1_dis}  and \eqref{plate2_dis} gives
\beq{7=22}
\begin{aligned}
       \hat V_1(\xi) &=   -\big(  q_+(\xi) - q_-(\xi) \big)\,\hat Y_1(\xi)
      + 4\pi p_0 \, \hat Y_1(k_0) \, \delta(\xi-k_0)  , 
\\
   \hat V_2(\xi) &=  -\big(  q_+(\xi) + q_-(\xi) \big)\,\hat Y_2(\xi) ,
\end{aligned}
\eeq
where 
\beq{qky2}
   q_\pm (\xi) = \frac{Z_{0\pm }}{d} \sum_{m=-\infty}^\infty    \big( \hat V_2(\xi + \frac{2\pi m}{d}) \pm \hat V_1(\xi + \frac{2\pi m}{d}) \big) ,
\eeq
with admittances
\beq{-32}
\hat Y_j (\xi) = \big\{ {\hat Z_{pj}(\xi)+ \hat Z_{fj}(\xi)}  \big\}^{-1}, \quad j=1,2,
\eeq
and plate impedances  
\beq{Zplate1}
\hat Z_{pj}(\xi) = \frac {D_j \xi^4 -m_j\omega^2}{-\ii \omega}, \quad j=1,2.
\eeq
The  latter are based on   the Kirchhoff plate theory.  Mindlin plate theory  is an alternative and arguably more accurate model, but  from related  work  \cite{BakhtiaryYekta2024} it is not expected to provide a noticeable difference. 

Noting that the functions $q_\pm$ are periodic, $ q_\pm(\xi) = q_\pm(\xi+\frac{2\pi m}{d}) $ for integer $m$, 
it follows from Eqs.\   \eqref{7=22} and \eqref{qky2} that
\beq{qky3}
\begin{pmatrix}
       q_+(\xi)  \\    q_-(\xi) 
\end{pmatrix}
=   \frac{4\pi p_0 \, \hat Y_1(k_0) }{ 1 + \big( \frac d{Z_{0+}} - \frac d{Z_{0-}} \big) \hat Z_0  (k_0) } \, 
\begin{pmatrix}
       \hat Z_+(k_0)  + 2  \hat Z_0 (k_0) \\  -  \hat Z_-(k_0)
\end{pmatrix}
\, \sum_{m=-\infty}^\infty 
    \delta \big(\xi-k_0 - \frac{2\pi m}{d} \big)  
\eeq 
where  
\beq{Zhat_k0}
   \begin{aligned}
 \hat Z_-(\xi) &=  \Big\{ \frac{d}{Z_{0-}} + 2 \hat S_1(\xi) \Big\}^{-1} ,
  \\
  \hat Z_+(\xi) &= \Big\{ \frac{d}{Z_{0+}} + 2 \hat S_2(\xi) \Big\}^{-1} ,
   \\
    \hat Z_0 (\xi) &=  \hat Z_+(\xi)\, \hat Z_-(\xi)\, \big(  \hat S_2(\xi) - \hat S_1(\xi) \big)
   \end{aligned}
\eeq 
and 
\beq{X}
 \hat S_j(\xi)  =  
 \sum_{m=-\infty}^\infty     \hat Y_j \big(\xi +  \frac{2\pi m}{d}\big)   , \ \ j=1, 2.
\eeq

\subsection{Reflected and transmitted waves}

Equations  \eqref{7=3}, \eqref{7=22} and  \eqref{qky3}  together yield the scattered pressure on either side, 
  \bal{55+}
 p_j(x,y) &= 2p_0 \,  \hat A_j (k_0)    \sum_{m=-\infty}^\infty 
 \hat  Z_{fj}\big( \xi_m \big)  
 \hat Y_j\big( \xi_m \big) e^{\ii \big( (-1)^j (k_{1x})_m\, x + \xi_m y \big)} 
 \notag \\ & 
-2p_0\,   \hat Z_{f1} (k_0) \hat Y_1(k_0)  \,  e^{k_1(-x \cos\theta_1 +y \sin \theta_1)} \, \delta_{j1} ,
\quad j = 1,2, 
 \eal
where
\beq{4+6}
\hat  A_1 (\xi) =  \frac{ \big(  \hat Z_+(\xi)+\hat Z_-(\xi) + 2  \hat Z_0(\xi) \big) }
{ 1 + \big( \frac d{Z_{0+}} - \frac d{Z_{0-}} \big)  \hat Z_0(\xi)   } \,  \hat Y_1(\xi),
 \qquad
 \hat  A_2 (\xi) =  \frac{   \hat Z_+(\xi)\hat Z_-(\xi) \,  \hat Y_1(\xi) }
{ \big( \frac d{Z_{0+}} - \frac d{Z_{0-}} \big)^{-1} +  \hat Z_0(\xi)  } ,
\eeq
and 
\beq{7=17}
 \xi_m = k_0 + 2\pi \frac md, 
\qquad
 (k_{jx})_m =  \sqrt{k_j^2 -  \xi_m^2 }
 \quad \text{ for} \quad m\in \mathbb{Z} .
\eeq
\ann{We assume that only the fundamental $m=0$ scattered modes propagate in air and water.   
All other  Bragg wavenumbers in the   $x$ direction, $ (k_{jx})_m $, $m\ne 0$, are positive imaginary, leading to evanescent acoustic fields. For normal incidence this requires that $k_2 < 2\pi/d$ or equivalently ${fd}/{c_a} < 1$.  The value of ${fd}/{c_a} $  does not exceed $0.2$ in  the numerical examples discussed below.
}

Total pressure in the incident water $(x<0)$ and the transmitted medium air $(x>0)$ follows from Eqs.\   \eqref{-44}  and \eqref{55+}  as 
\beq{Ref1Coef}
    p(x,y) =
    \begin{cases} p_0 e^{ \ii k_1(x \cos\theta_1 +y \sin \theta_1)} + p_0 R(\theta_1)\, e^{\ii k_1(-x \cos\theta_1 +y \sin \theta_1)}
    +p_{1ev}(x,y) , & x<0, 
    \\
    p_0 \, T(\theta_2)\, e^{\ii k_2(x \cos\theta_2 +y \sin \theta_2)}
    +p_{2ev}(x,y),  & x>0, 
   \end{cases}
\eeq
where
\beq{RefCoef}
\begin{aligned}
   R(\theta_1) =& R_1(\theta_1) +  \big( 1-R_1(\theta_1) \big)\, 
    \hat A_1 (k_0) ,
    \\
     T(\theta_2) =&   \big( 1-R_2(\theta_2) \big)\, 
    \hat A_2 (k_0)  ,
\end{aligned}
\eeq
 $R_1$ and  $R_2$ are the reflection coefficient for plane wave incidence on the plates,   
\beq{Ref1Coef+}
    R_j(\theta_j) = \frac{\hat Z_{pj}(k_0) - \hat Z_{fj}(k_0)}{\hat Z_{pj}(k_0) + \hat  Z_{fj}(k_0)} ,
\eeq
and the evanescent, or near, fields,  are
\beq{pev1}
    p_{jev}(x,y) = 2p_0 \,  \hat A_j (k_0)  
  \, 
    \sum_{m \neq 0}
    \hat  Z_{fj}\big( \xi_m \big)  \hat Y_j\big( \xi_m \big)  
     e^{\ii \big( (k_{jx})_m\, |x|  +\xi_m y  \big) } .
\eeq
Energy conservation requires that
\beq{342}
|R|^2 + \frac{Z_w \sec\theta_1} {Z_a \sec\theta_2}\, |T|^2 = 1 .
\eeq

Note that the reflection coefficient  can be expressed in the alternative form
\beq{7=73}
   R(\theta_1) =\frac{\hat Z_{p1}'(k_0) - \hat Z_{f1}(k_0)} {\hat Z_{p1}'(k_0) + \hat  Z_{f1}(k_0)} 
\eeq
suggestive of reflection from a plate with impedance
\beq{7=74}
 \hat Z_{p1}'(\xi) = \hat Z_{p1}(\xi) + \frac{  \hat  A_1(\xi) }{ \big( 1-  \hat  A_1(\xi)\big) \hat  Y_1(\xi) }.
\eeq


\section{Conditions for full transmission}   \label{sec5}

Total transmsission  corresponds to $R=0$, implying two conditions for the real and imaginary parts.  In order to understand these conditions  we consider the case of rigid and massless ribs.   This has little effect on the full solution, as we will see, but it significantly simplifies the algebra, allowing us to find the necessary constraints on the system parameters.   

\subsection{Frequency for total transmission}  \label{sec5a}  
In the  limit that the ribs are  rigid, $1/Z_{0-}\to 0$, and massless,  $Z_{0+} \to 0$: 
$\hat  A_1 (\xi) = \hat  A_2 (\xi) =  {\hat  Y_1 (\xi) }/\big( \hat  S_1 (\xi) +\hat  S_2 (\xi)  \big) $  and 
the reflection coefficient takes the form
\beq{7073}
   R(\theta_1) =\frac{  R_1(\theta_1)\, \Gamma (k_0)}{\hat  S_1 (k_0) +\hat  S_2 (k_0)  }
\eeq
where 
\beq{7-3}
   \Gamma (k_0)=    \hat S_1' (k_0) +\hat  S_2' (k_0)
    +  \frac 1{\hat Z_{p1}(k_0) - \hat Z_{f1}(k_0)} +   \frac 1{\hat Z_{p2}(k_0) + \hat Z_{f2}(k_0)}
\eeq
with  $ \hat S_j'(\xi)  = \hat S_j (\xi) -  \hat Y_j \big(\xi)$, $j=1,2$.
Total transmission corresponds to zero reflection, and we therefore look at the conditions required to make $ \Gamma$ and hence $R$  vanish.   We  consider normal incidence, $k_0 = 0$.  

Under these circumstances  $ \hat S_1' (k_0) $ and $\hat  S_2' (k_0)$ are imaginary. Setting the real part of $ \Gamma (0)$ in \eqref{7-3} to zero 
yields the transmission frequency 
 $\omega_0$: 
 \beq{429}
\omega_0^2 = \frac{Z_wZ_a \, (Z_w-Z_a)}{ Z_w m_2^2 - Z_a m_1^2} .
 \eeq
 Equivalently, 
  \beq{428}
\omega_0 = \frac{\sqrt{Z_wZ_a}}{m_2} \,  \bigg( \frac{1 - \epsilon } {1 - \epsilon\, m_1^2/m_2^2}
\bigg)^{1/2} 
 \eeq
 where $ \epsilon \ll 1$ is defined in \eqref{3+0}. 
 This provides the remarkable simplification
 \beq{440}
 \omega_0 \approx \frac{Z_e}{m_2} \quad \text{where}\quad Z_e \equiv \sqrt{Z_wZ_a},
 \eeq
 which  is a very accurate approximation to Eq.\ \eqref{428} on account of the smallness of 
 $ \epsilon$.

 An alternative and simpler method is presented in Appendix \ref{appb} for finding the frequency of full transmission, Eq.\  \eqref{429}. 

\subsection{Optimal rib spacing  }

Setting the imaginary part of  $\Gamma (k_0)$ of \eqref{7-3}  to zero 
at   $\omega = \omega_0$ given by \eqref{440}, with $k_0=0$ yields 
 \beq{top}
 2
 \sum_{n=1}^\infty \Big \{ 
 \Big( \hat D_1 n^4 - \frac{m_1}{m_2} - \frac{\rho_w d}{m_2 2\pi n}      \Big)^{-1}
 +  \Big( \hat D_2 n^4 -1    \Big)^{-1}
 \Big\}   \approx 1
 \eeq
 where $\hat D_j =  \frac{m_2 (2\pi)^4}{Z_e^2 d^4} D_j$, 
  $j=1,2$ and the  approximations  $\epsilon \ll1$ and  $\rho_a d/m_2 \ll 1$ have been used.    Equation \eqref{top} determines $d$ for chosen  $h_1$, or {\it vice versa}.  For instance, if $\omega_0$ and $h_1$ are chosen, along with the plate materials (e.g.\ both aluminum), then Eq.\ \eqref{440} defines $h_2$ and \eqref{top} determines $d$.  

 Equation \eqref{top} has a close  connection with the quasistatic stiffness of the two-plate flex-layer  \cite{BakhtiaryYekta2024} as we now explain.  Ignoring the inertial terms, which is consistent with the quasistatic limit, \eqref{top} becomes 
  \beq{6-7}
 2
 \sum_{n=1}^\infty \Big \{ 
 \Big( \hat D_1 n^4  \Big)^{-1}
 +  \Big( \hat D_2 n^4    \Big)^{-1}
 \Big\}   \approx 1
 \quad \Rightarrow \quad d\approx  d_0
 \eeq
where 
  \beq{059}
d_0^4 = \frac{720 }{ Z_\text{e} \, \omega_0} \Big( \frac 1{D_1} +   \frac 1{D_2}  \Big)^{-1} 
\eeq
and the identity 
 $ \sum_{n=1}^\infty \frac 1{n^4} =   \frac{\pi^4}{90 } $ has been used.   
The relation \eqref{6-7}$_2$ for $d$ can be understood in terms of the effective  quasistatic stiffness $\kappa_\text{eff}$ of the flex-layer introduced in  \cite{BakhtiaryYekta2024} for a symmetric plate system.   In the present case the plates are different and we need to take the  flexural stiffness of the plates in series, i.e.\   
$\kappa_\text{eff}  = ( \kappa_1^{-1} + \kappa_2^{-1} )^{-1} $  where $\kappa_j = 720 D_j/d^4$  \cite{BakhtiaryYekta2024}.
 The connection with \eqref{6-7}$_2$ follows from the resonance condition  $\kappa_\text{eff} = m\omega_0^2$   for  effective mass $m=m_2$.    
Together   with  \eqref{440} this yields  $\kappa_\text{eff} = Z_e \omega_0$
which then  implies the relation for   $d$ according to \eqref{6-7}$_2$.

Assuming $h_1$ is chosen, then $d$ follows approximately from the  estimate $d_0$ of Eq.\  \eqref{059}.    We use $d_0$ as an initial estimate for the solution of  the nonlinear equation \eqref{top}.

 \section{Numerical demonstrations and discussion} \label{sec6} 

Illustrative examples are presented  based on  the derived equations in sections \ref{sec4} and \ref{sec5}.  In all cases both plates are Aluminum ($\rho_s = 2,700 $  kg/m$^3$, $E = 70$ GPa, $\nu=0.334$), and the thickness of plate 1 is $h_1 = 1$ mm. 
The theoretical equations are verified by numerical simulation, using  COMSOL with the multi-physics options set to "2-D Acoustic-Solid Interaction, Frequency Domain"  using  "Perfectly Matched Boundary" and "Periodic Condition" on the model  boundaries.  
The ribs are assumed to be infinitely stiff, with a length of 1 cm and a thickness of 1 mm.
\subsection{Full transmission } \label{sec6a} 
Full transmission at  a given frequency $f_0$ requires that the lengths  $h_2$ and $d$ assume optimal values according to Eqs.\ \eqref{440} and \eqref{top}.  We proceed by choosing the transmission frequency $f_0$, then  find  $h_2 \approx Z_e/(2\pi f_0 \rho_{s2})$  from Eq.\ \eqref{440} and subsequently  use this value to  find $d$ from  Eq.\  \eqref{top}, with the initial guess $d=d_0$  of Eq.\  \eqref{059}.
\begin{figure}
    \centering
 \includegraphics[width=0.9\textwidth]{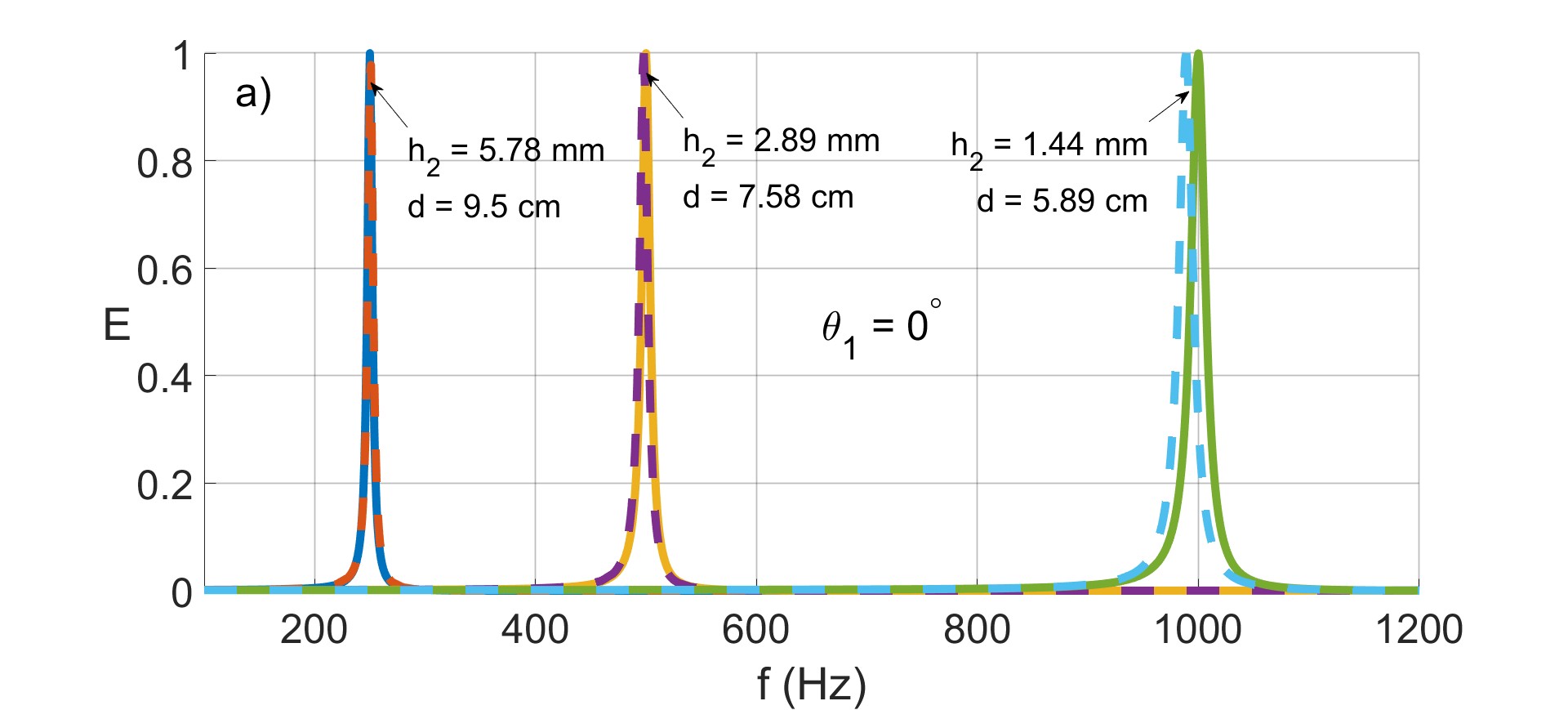}  \\
\includegraphics[width=0.9\textwidth]{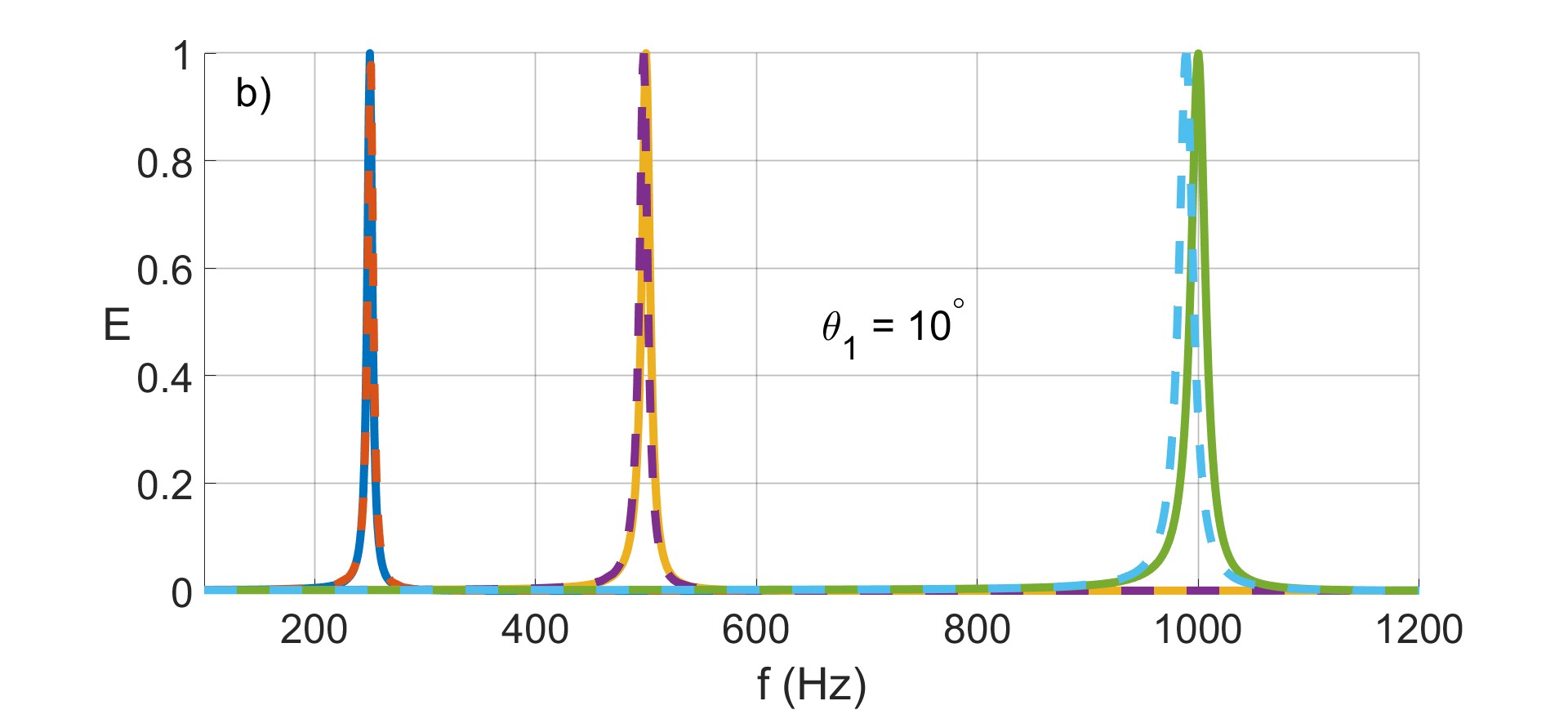} \\
 \includegraphics[width=0.9\textwidth]{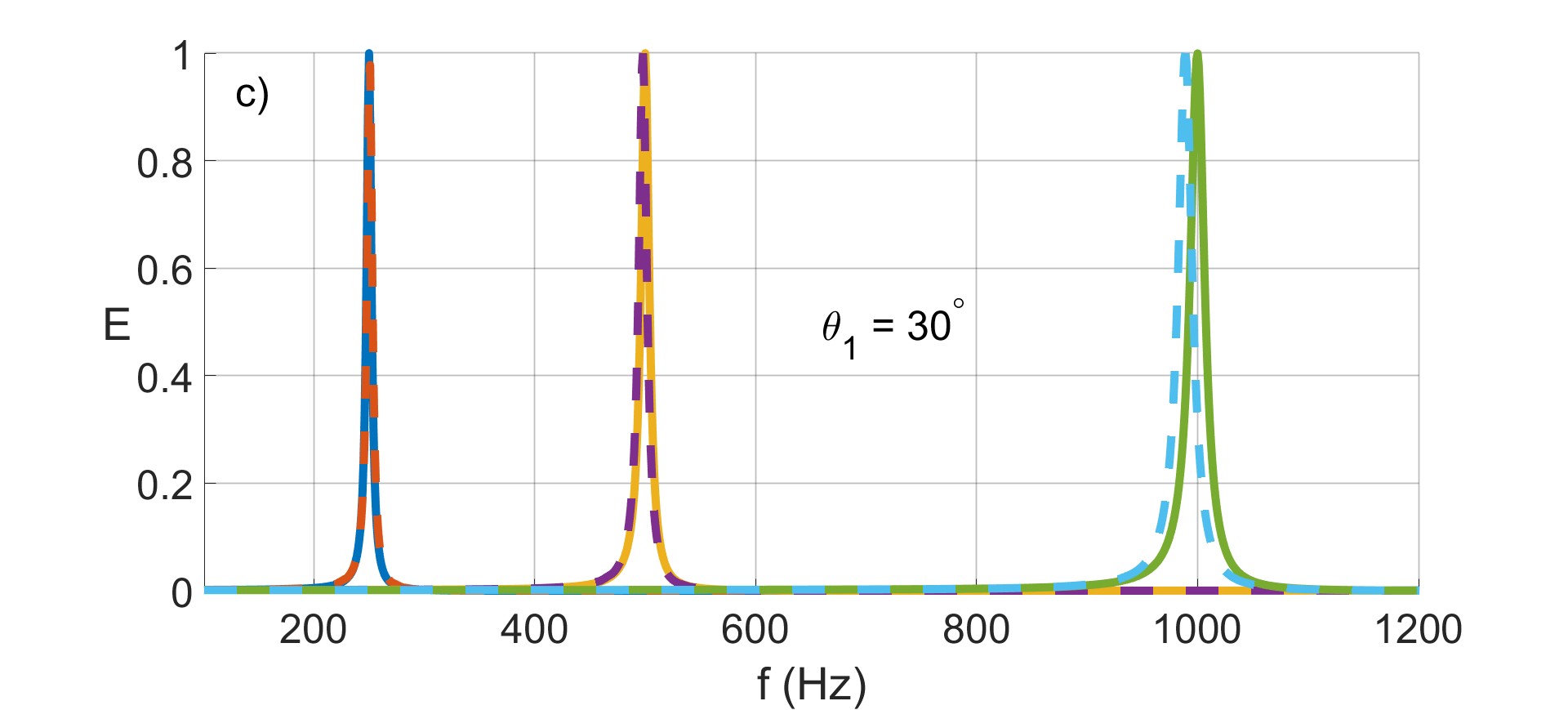} 
    \caption{The transmitted acoustic energy E vs frequency for the flex-layer model based on optimal values obtained  for a) $\theta_1 = 0 ^\circ$, b) $\theta_1 = 10 ^\circ $ and  c) $\theta_1 = 30 ^\circ $. Dashed lines are obtained using COMSOL. In all cases $h_1 = 1$ mm.}
    \label{theta0_E2}
\end{figure}

We consider three different flex-layers with parameters based on the optimal values of $d$ and $h_2$ for normal incidence ($\theta_1 = 0^\circ$) at transmission frequencies $f_0 = 250$ Hz ($d = 9.5$ cm and $h_2 = 5.78$ mm), $f_0 = 500$ Hz ($d = 7.58$ cm and $h_2 = 2.89$ mm), and $f_0 = 1000$ Hz ($d = 5.89$ cm and $h_2 = 1.44$ mm).
Figure  \ref{theta0_E2}  shows the transmitted acoustic energy E\, $ = \frac{z_w \sec\theta_1} {z_a \sec\theta_2}\, |T|^2$ for both normal incidence and for incident angles  $\theta_1 = 10^\circ$ and $\theta_1 = 30 ^\circ $.   
It is evident from Fig.\  \ref{theta0_E2} that the optimal $h_2$ and $d$  for normal incidence also work for oblique incidence, and that the full-transmission frequency is independent of $\theta_1$ for a given optimized flex-layer. 
 \textcolor{black}{Simulations based on the theory assumes that the rib mass and thickness   are negligible. These assumptions can lead to slight deviations between the simulation and COMSOL results at higher frequencies,  as for example, in Fig.\  \ref{theta0_E2} for the 1000-Hz case.}

\begin{figure}
    \centering
    \includegraphics[width=0.9\textwidth]{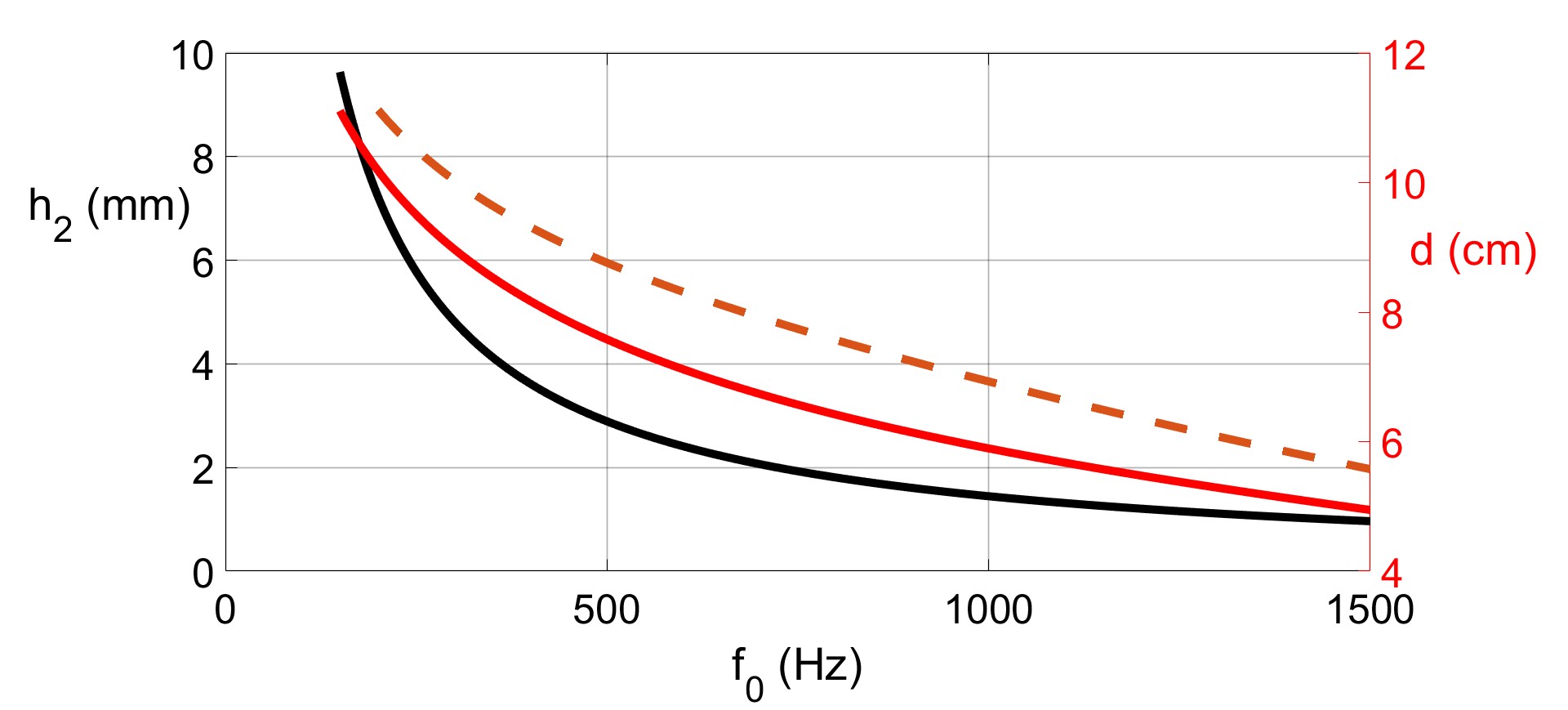}
    \caption{Optimal values for $h_2$ and $d$ vs $f_0$ for normal incidence. $h_1 = 1$ mm. The dashed  curve is the approximation $d_0$ of Eq.\ \eqref{059}.}
    \label{fmax_h2_d}
\end{figure}

Figure \ref{fmax_h2_d} illustrates the relation between $f_0$, $h_2$, and $d$ for a range of transmission frequencies.   The values of the approximate spacing $d_0$ is also shown, indicating that it is  an overestimate of the optimal spacing for the parameter range  considered. 
The optimal $h_2$ and $d$ in  Fig.\ \ref{fmax_h2_d} are calculated for normal incidence $\theta_1 = 0^\circ$.   Based on the results of  Fig.\  \ref{theta0_E2}  we can safely surmise that the same optimal values apply for 
 $\theta_1 \ne 0^\circ$.

\subsection{Bandwidth and Q-factor} \label{sec6b} 

\begin{figure}
    \centering
 \subfigure[Effect of varying $h_1$ with  $E_1 =  70$ GPa]{ \includegraphics[width=0.9\textwidth]{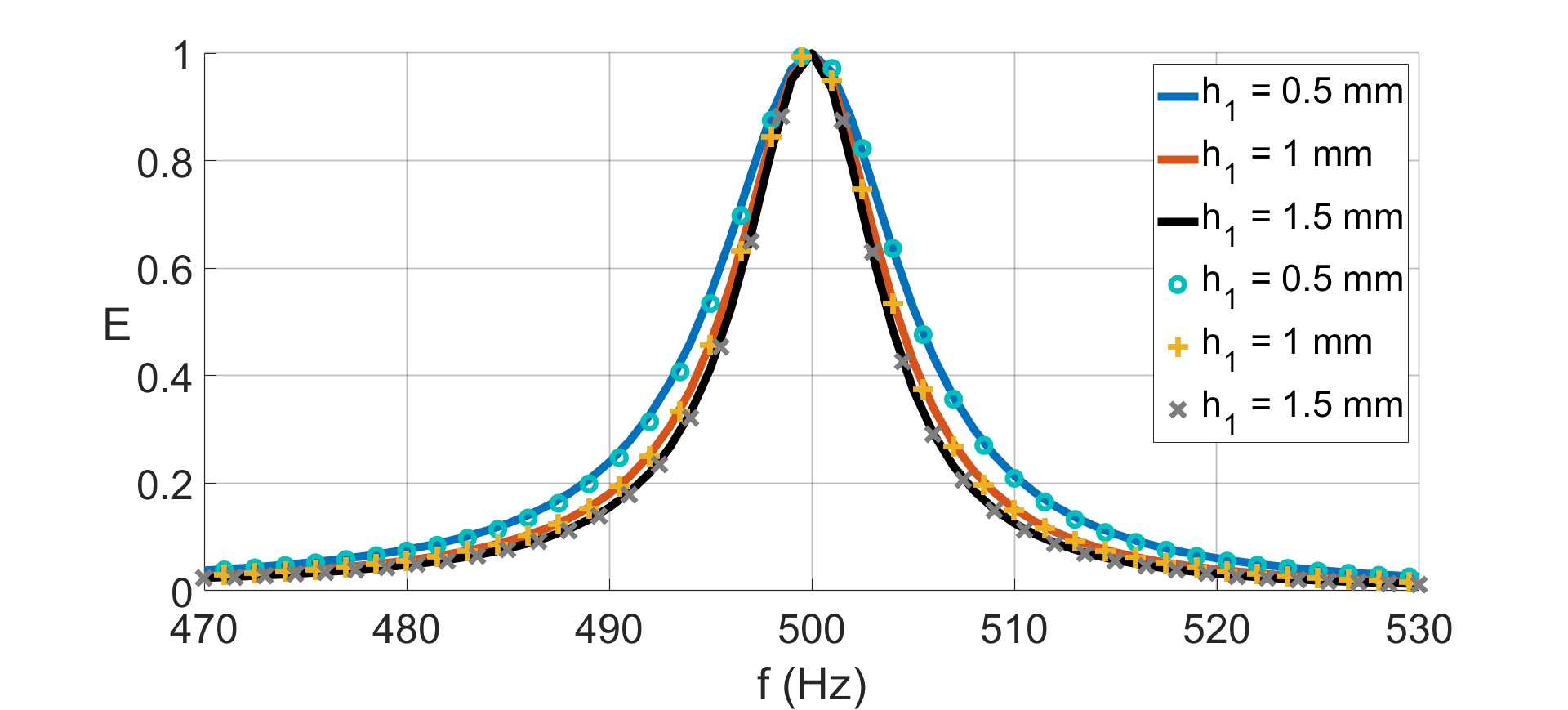} }
  \subfigure[Effect of varying $E_1$ with $h_1 = 1$  mm]{ \includegraphics[width=0.9\textwidth]{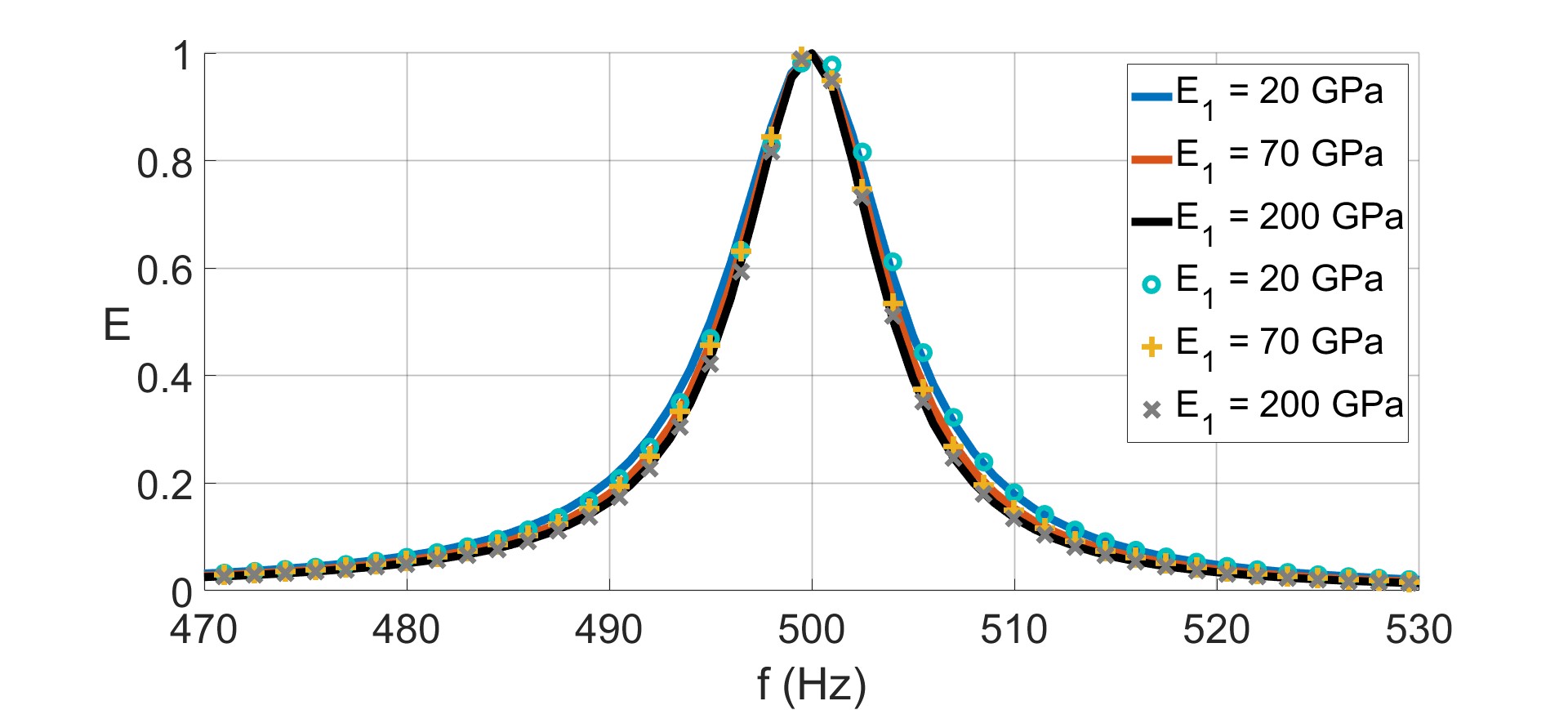} }
    \caption{ The effect of varying  bending stiffness parameters for plate 1.  The solid curves are theory and the symbolic shapes $\circ$ , $+$, and $\times$ are COMSOL results.}
    \label{hEeffect}
\end{figure}

We consider the effect of some system parameters on the bandwidth of the acoustic transmissivity. The parametric studies are conducted in such a way that if the same resonant frequency is desired, then changing the parameters of the first plate also changes $d$ to maintain the equivalent bending stiffness of the first plate.  We focus on full transmission at $f_0 = 500$ Hz. Thus,  for the case that $h_1 = 1$ mm, the optimal value for $d$ is 7.58 cm, while for $h_1 = 0.5$ and $1.5$ mm, the optimal values for $d$ are $4.75$ cm and $9.86$ cm, respectively). The results demonstrate that $h_1$ has an inverse relationship with the bandwidth. Consequently, by decreasing $h_1$, the bandwidth increases, as depicted in Fig.\ \ref{hEeffect}(a). The next parameter we investigated to observe its effect on the bandwidth is the Young's modulus of the first plate, $E_1$. As shown in Fig.\  \ref{hEeffect}(b), $E_1$ exhibits an inverse relationship with the bandwidth (for the case that $E_1 = 70$ GPa, the optimal value for $d$ is 7.58 cm, while for $E_1 = 20$ and $200$ GPa, the optimal values for $d$ are 5.69 cm and 9.58 cm, respectively). By comparing Figs.\  \ref{hEeffect}(a) and \ref{hEeffect}(b), it is evident that the effect of $h_1$ on the bandwidth is stronger than $E_1$. 

\begin{figure}
    \centering
    \includegraphics[width=0.9\textwidth]{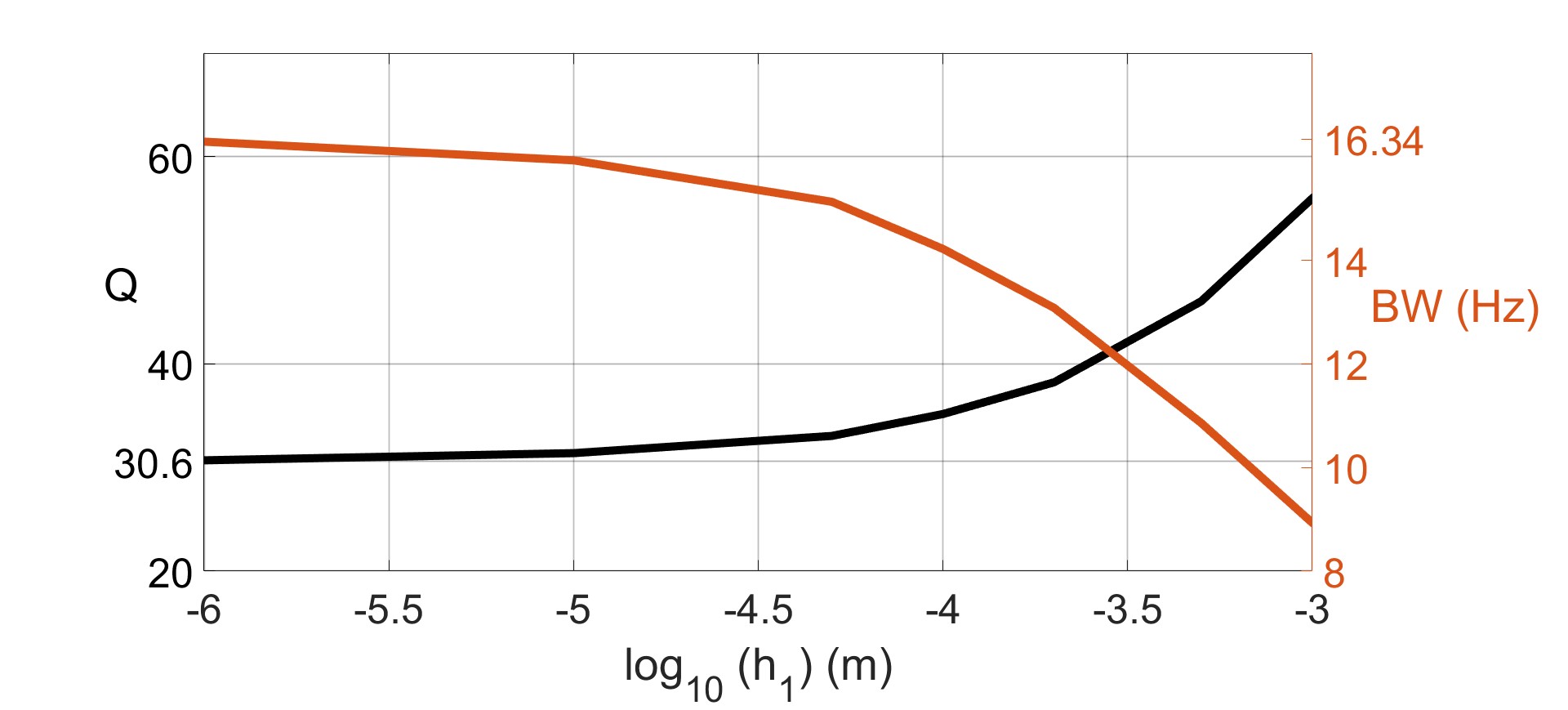}
    \caption{The effect of varying $h_1$  on the Q-factor and BW.}
    \label{h1effect_Q_BW}
\end{figure}

Finally,  the frequency bandwidth (BW) and   Q-factor for the acoustic transmission are studied, where Q$ = \frac{f_0}{\Delta f}$ with $\Delta f$ equal to the BW at E\, $ = \frac 12$.  As illustrated in Fig.\ \ref{h1effect_Q_BW}, the Q-factor decreases with decreasing $h_1$, converging to $\frac{1}{2 \sqrt \epsilon} \approx$ 30.6, where $\epsilon = {Z_a}/{Z_w}$. From the results in Figs. \ref{hEeffect}(a) and \ref{h1effect_Q_BW}, it is observed that by decreasing $h_1$,  the result  converges  to the simple model   of  Fig.\ \ref{fig2}, where the Q-factor and bandwidth were found to be $30.6$ and $16.34$ Hz, respectively.

\subsection{Motion of the plates}  \label{sec6c} 

In order to further understand the mechanics at play in  the full transmission effect it is instructive to consider the motion of plates 1 and 2 facing  water and air, respectively.  At total transmission the energy fluxes of the incident and transmitted waves are equal.  Assuming 1D propagation (as in the model of Section \ref{sec2}), implies $p_1v_1 = p_2 v_2$.  Using the plane wave relations 
$p_1 = Z_w v_1$ and $p_2 = Z_a v_2$ it follows that $v_1 = \pm \sqrt{\epsilon}\, v_2$ with $\epsilon$ defined in  \eqref{3+0}.    We now discuss whether or not  this relation is reflected in the numerical simulations.  The short summary is that it is approximately, but in an averaged sense.   The longer story requires some explanation.

\begin{figure}
    \centering
    \subfigure[Plate 1, water side]{\includegraphics[width=0.49\textwidth]{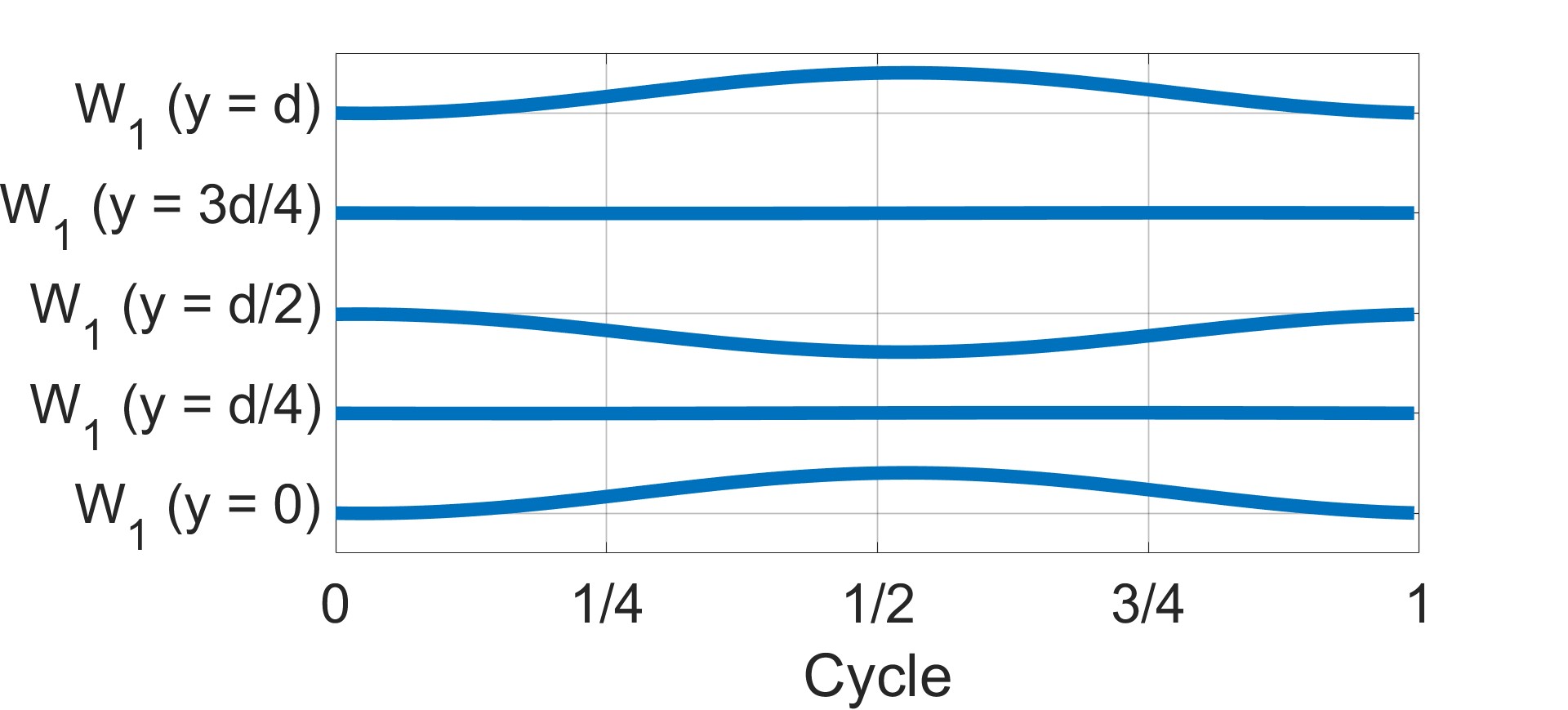}}
    \subfigure[Plate 2, air side]{\includegraphics[width=0.49\textwidth]{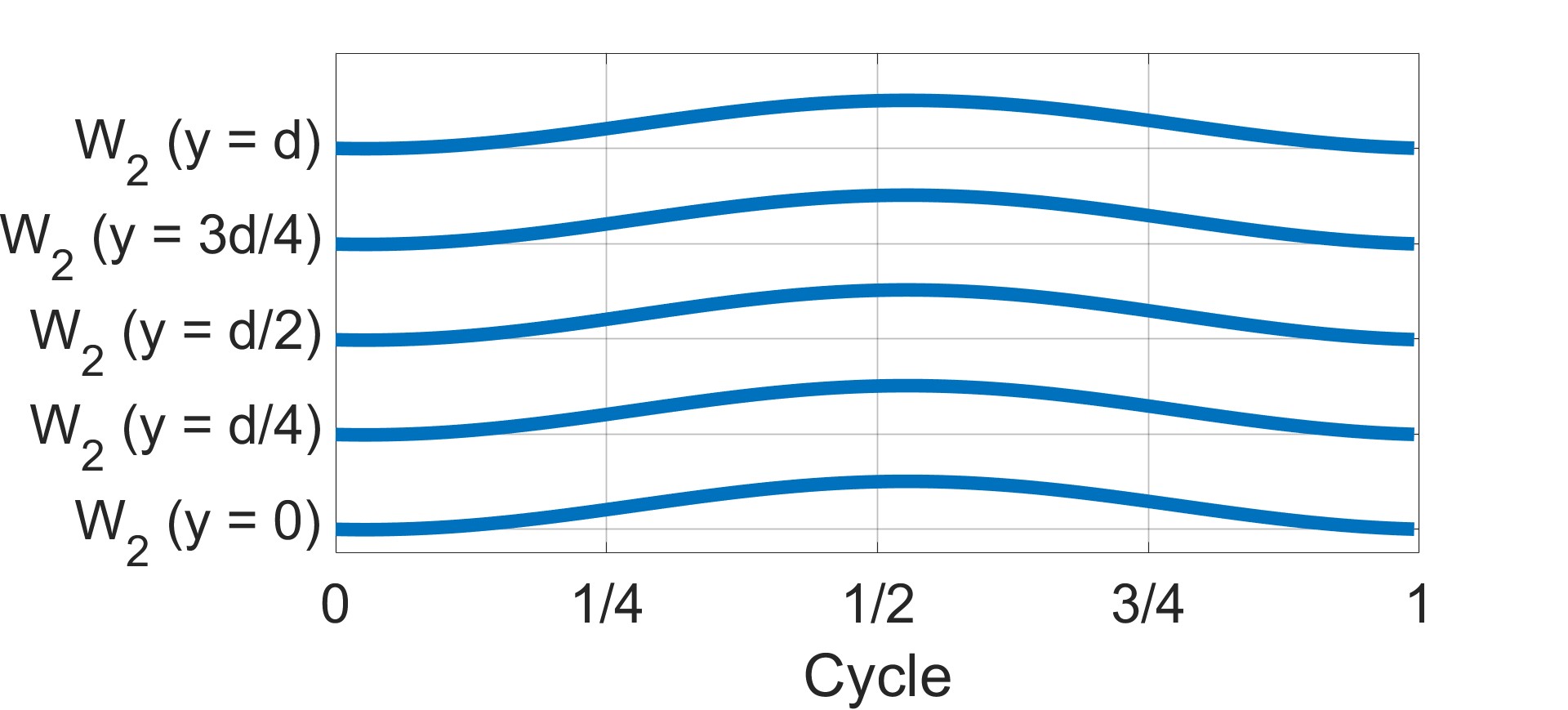}}
    \caption{The plate displacements $w_1 (t)$ and $w_2 (t)$ over a cycle, $0\le f_0 t \le 1$ for $f_0 = 500$ Hz.   The data is from the video 
\href{https://drive.google.com/file/d/1_NO-cwa2qGesVoLGW8-T1Ay3v3nVykby/view?usp=drive_link}{ Video 2, $f_0 = 500$ Hz}. }
    \label{W1_W2_cycle}
\end{figure}

Figure \ref{W1_W2_cycle} shows the plate displacements over one cycle for transmission frequency $f_0=500$ Hz. It is clear  that plate 2 on the air side moves like a plane wave, but the same is not true for plate 1.   The dramatic difference in the  plate motions is better appreciated from the associated videos for three transmission frequencies: 
\href{https://drive.google.com/file/d/1hwTII7HsRHWr4vbilIWEuffNaPOFEKSy/view?usp=drive_link}{ Video 1, $f_0 = 250$ Hz},
\href{https://drive.google.com/file/d/1_NO-cwa2qGesVoLGW8-T1Ay3v3nVykby/view?usp=drive_link}{ Video 2, $f_0 = 500$ Hz}, and
\href{https://drive.google.com/file/d/1hXlspb9I8zyB1HLhrAJKYhKuCblfUlr7/view?usp=drive_link}{Video 3, $f_0 = 1000$ Hz}.  To better  understand  the dramatically different plate dynamics, we refer to  Appendix \ref{appb}  which derives an alternative and simpler method for finding the frequency of full transmission, Eq.\  \eqref{429}.   The method uses spatial averages of the plate displacements, $\bar w_1$ and $\bar w_2$.  In addition to the frequency condition Eq.\  \eqref{429} it also follows from 
Eq.\  \eqref{3-45} that the average motion of  plate 1 small in comparison to that of plate 2, i.e. \ 
$ \bar w_1\approx -\ii \sqrt{\epsilon}\, \bar w_2 $.

We therefore have different expectations: 
$ \bar w_1\approx \eta \sqrt{\epsilon}\, \bar w_2 $
where $\eta$ can be $-\ii$, $+1$ or $-1$.
We find from simulation (COMSOL) that
$\frac { \bar w_1 }{   \sqrt{\epsilon}\, \bar w_2 } $ takes the values 
1.319 $e^{-0.53\, \ii \frac{\pi}{2}}$, 1.65 $e^{-0.36\, \ii \frac{\pi}{2} }$, 1.52 $e^{-0.445\, \ii \frac{\pi}{2} }$
for $f_0= 250$ Hz, $500$ Hz and $1000$ Hz, respectively.  We conclude that none of the above are correct, although the magnitude is close, $ |\bar w_1|\approx  \sqrt{\epsilon}\, |\bar w_2| $, indicating very little average motion of plate 1. 
The difference can be ascribed to the assumptions  used in Appendix \ref{appb}, specifically that the pressure and velocity on plate 1 are related by the plane wave impedance relation $p_1 = Z_w  \bar v_1$, which is clearly not the case.  Near-field evanescent effects are  important at plate 1 but are not included in the analysis of Appendix \ref{appb}.

\begin{figure}
    \centering
    \subfigure[$f_0 = 250 $ Hz]{\includegraphics[width=0.49\textwidth]{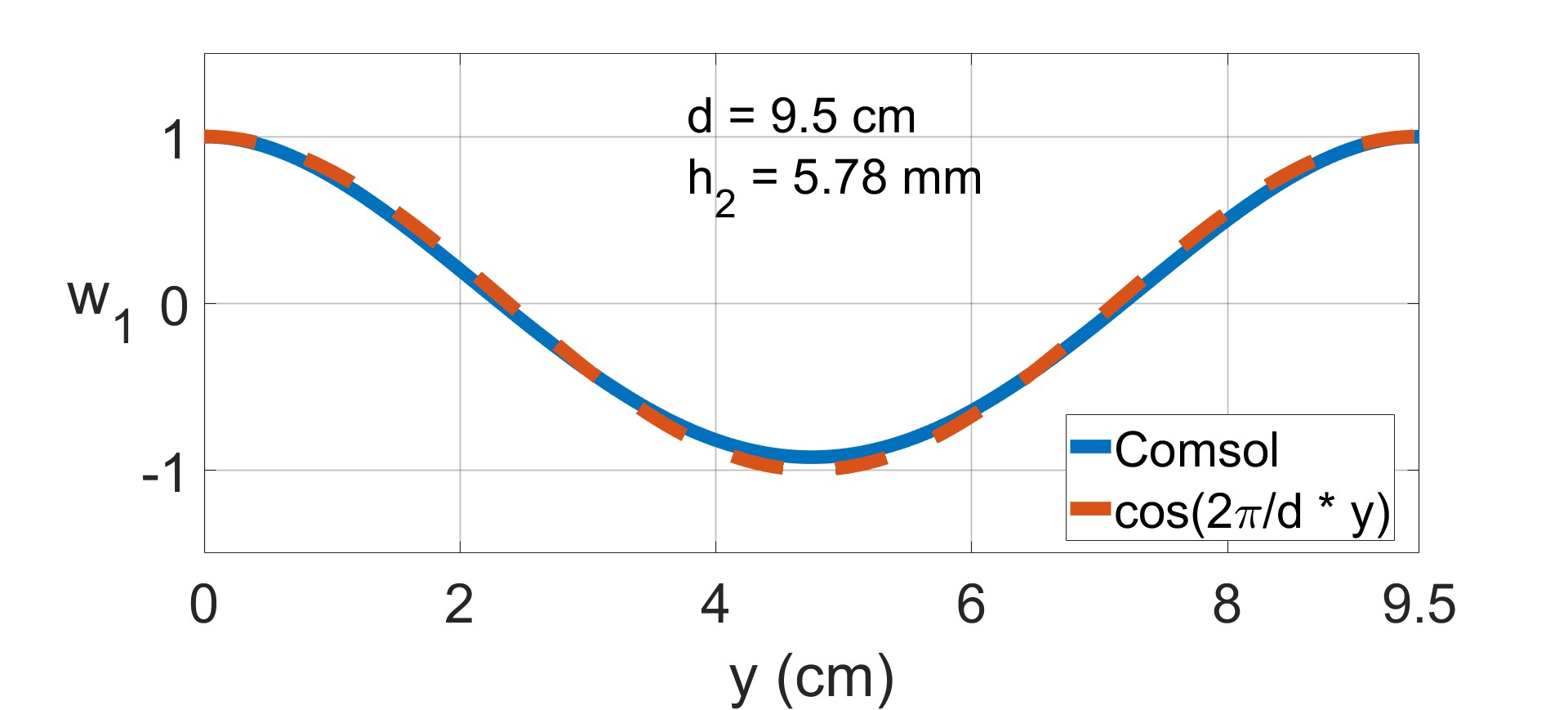}}
    \subfigure[$f_0 = 500 $ Hz]{\includegraphics[width=0.49\textwidth]{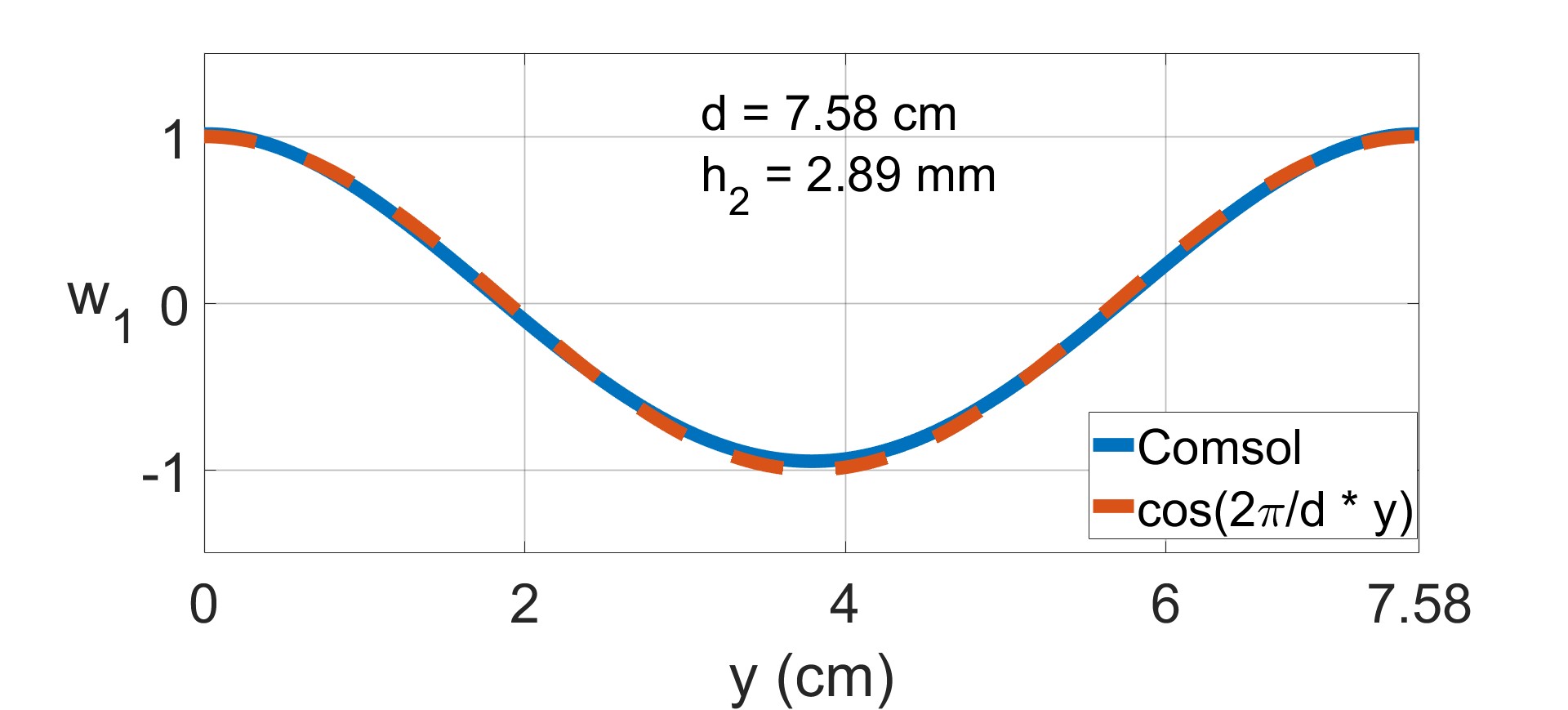}}
    \subfigure[$f_0 = 1000 $ Hz]{\includegraphics[width=0.49\textwidth]{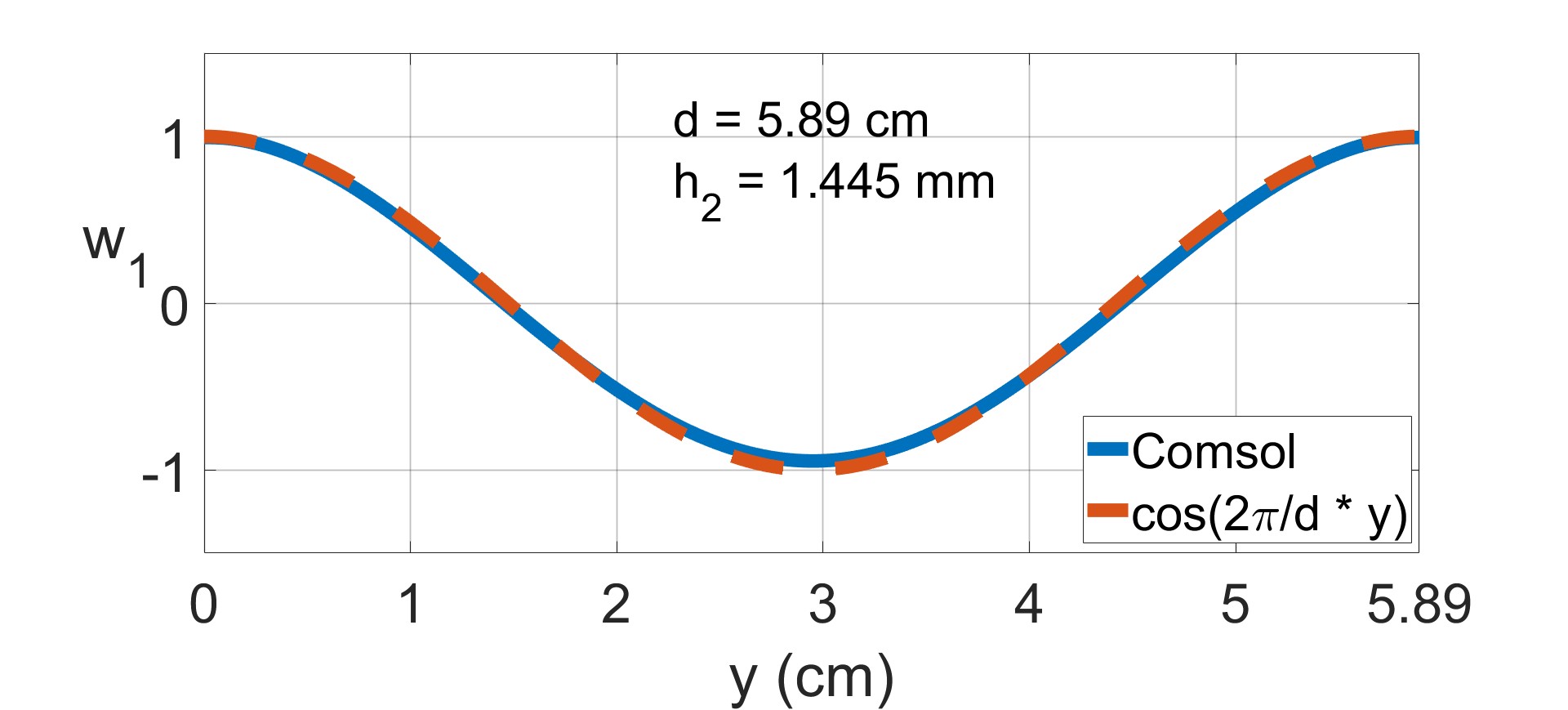}}
    \caption{The mode shape of plate 1 facing water at full transmission for three transmission frequencies:   $250 $ Hz, $ 500$ Hz   and $ 1000$ Hz.}
    \label{ShapeMode}
\end{figure}

Finally, we note that the mode shape of   plate 1 on the water side can be accurately modeled if we ignore the effect of fluid loading and consider the plate equation only.  The mode must be symmetric in $y$ with zero slope at $y = \pm \frac d2$, and hence
\beq{wasym}
w_1(y) = C_1 \big[ \sinh \big( \frac{\beta d}2 \big) \cos(\beta y) + \sin \big( \frac{\beta d}2 \big)  \cosh(\beta y) \big]
\eeq
where $\beta^4 = \omega_0^2 m_1/D_1$.  
We assume that the shear force at  $y = \pm \frac d2$ is approximately zero, i.e.\   $D_1 w_1 ''' \big( \pm \frac d2 \big)  \approx 0$,  
implying $    \sin \big( \frac{\beta d}2 \big) \approx 0$.
Taking the first non trivial solution,  
$\beta \approx \frac{2\pi}{d}$, yields the simple mode shape
\beq{3+5}
w_1(y) \approx A_1 \cos \big(\frac{2\pi}{d} y \big)
\eeq
for some $A_1$.   Figure  \ref{ShapeMode}  indicates remarkable  agreement between the light fluid loading model of Eq.\ \eqref{3+5} and COMSOL simulations of the water-facing plate. 

\subsection{Effect of air between the plates} \label{sec6d} 

We consider how the assumption of vacuum between the plates compares with the more realistic scenario of entrained air.  The presence of air introduces an equivalent spring between the plates with stiffness $\kappa_a$ similar to that in Section \ref{sec2}.  In this case $\kappa_a = \rho_a c_a^2 /L_r$ where $L_r$ is the rib length, i.e.\ the distance separating the plates. Also, based upon the previous results for the plate motions, it expected that the relative displacement $\bar w_2- \bar w_1$ is well approximated by $\bar w_2$.  Therefore, as a first  approximation we assume that the effect of the air is to change the effective acceleration  of plate 2 from $-m_2\omega^2 w_2$ to approximately 
 $-(m_2\omega^2 - \kappa_a)w_2$.  Using this in the derivation of \eqref{429} from \eqref{7073} yields  a  transmission frequency $\omega$ greater than $ \omega_0$ of Eq.\ \eqref{44-0} which  satisfies 
 \beq{022}
 m_2^2\, \omega^4 - (Z_wZ_a + 2 m_2 \kappa_a)\, \omega^2 + \kappa_a^2 \approx 0 . 
 \eeq
 Hence
 \beq{888}
 \frac{\omega}{\omega_0} 
 \approx \frac 12 +  \sqrt{\frac 14  + \frac{\sqrt{\epsilon} } {\lambda}}  
  \quad \text{where} \quad \lambda = \frac{ \omega_0 L_r}{c_a}.
 \eeq
 The dependence on transmission frequency and rib length combine in the single non-dimensional parameter $\lambda$.

\begin{figure}
    \centering
    \subfigure[ $f_0 = 150$ Hz]{\includegraphics[width=0.49\textwidth]{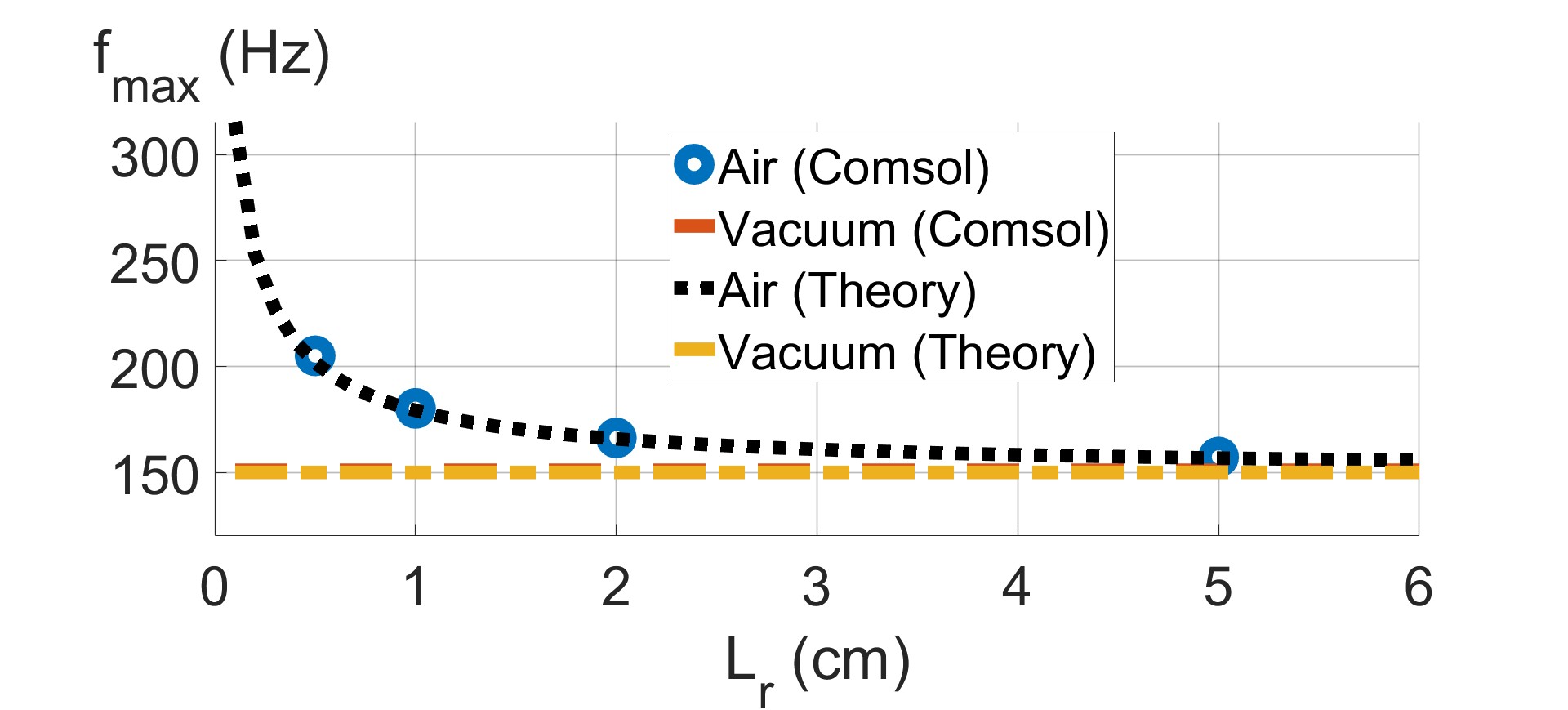}} 
    \subfigure[ $f_0 = 500$ Hz]{\includegraphics[width=0.49\textwidth]{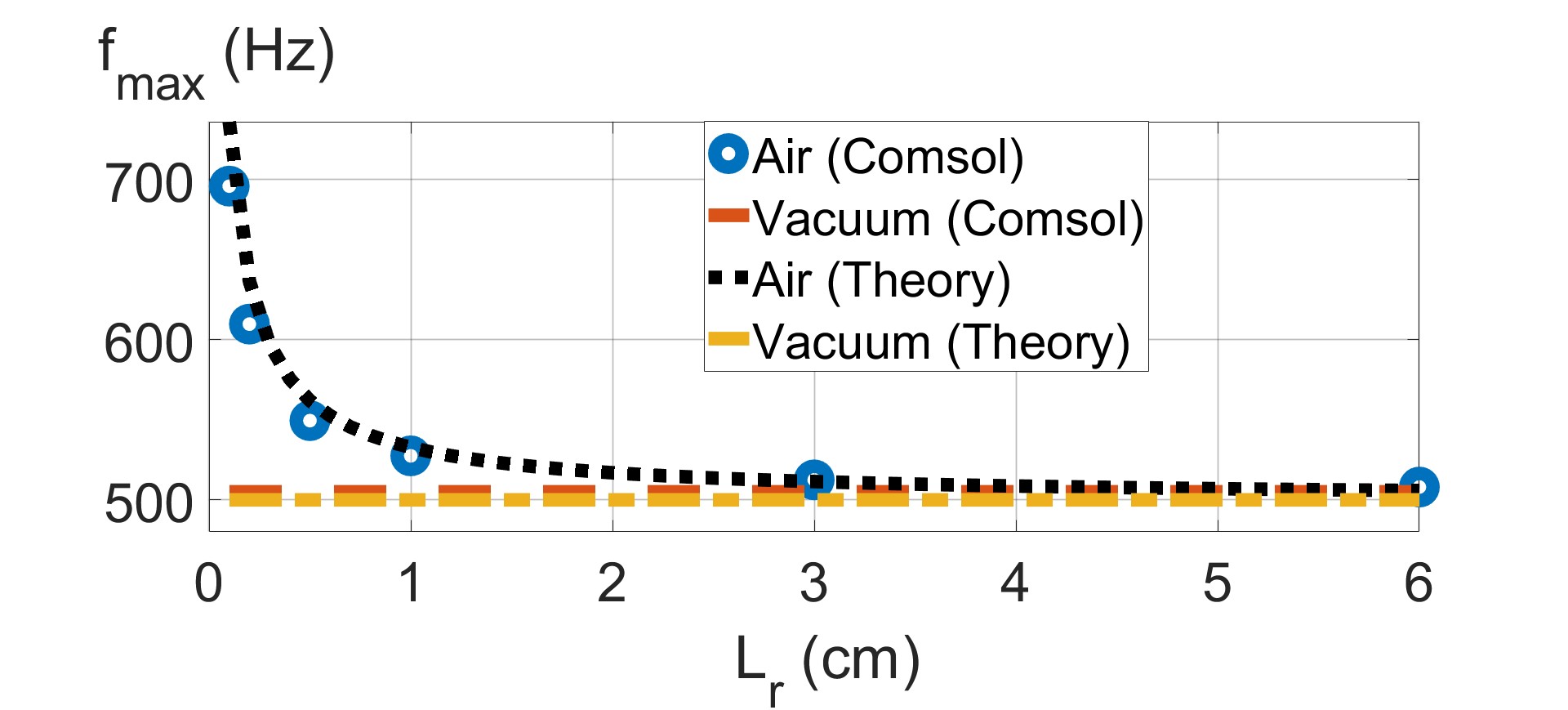}}
    \subfigure[ $f_0 = 1000$ Hz]{\includegraphics[width=0.49\textwidth]{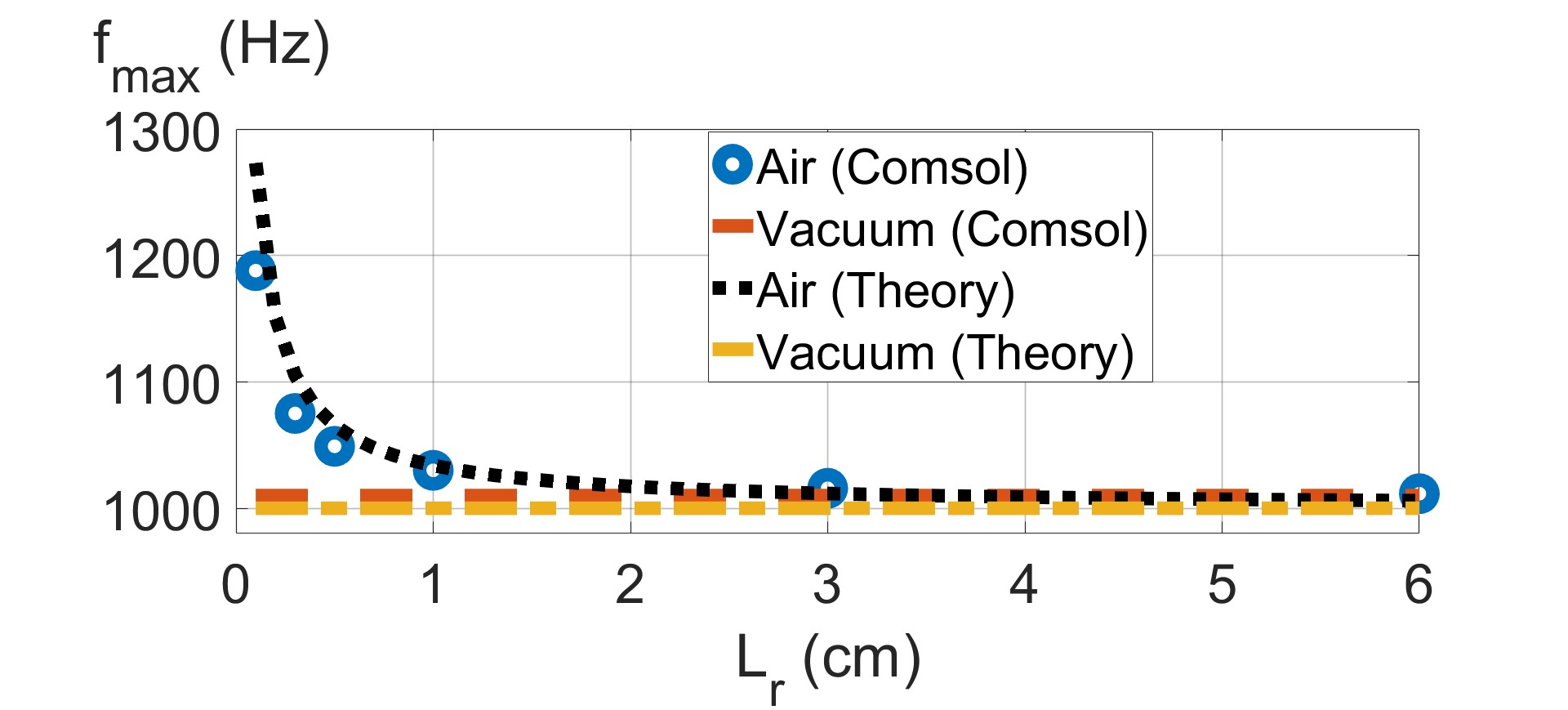}} 
    \caption{The effect of air between the plates as a function of the rib length $L_r$ for three full transmission frequencies: (a) $f_0 = 150$ Hz,  (b) $f_0 = 500$ Hz, (c) $f_0 = 1000$ Hz.}
    \label{fmax_air_vacuum}
\end{figure}

Figure \ref{fmax_air_vacuum} demonstrates that Eq.\ \eqref{888} accurately predicts the effect of the presence of air on the  flex-layer transmission  frequency. In particular, the  frequency shows an inverse relationship with the length of the ribs, indicating that shorter rib lengths result in stronger spring characteristics of the air.  The resonant frequency increases as the air volume decreases and $\kappa_a$ increases.  However, by taking $L_r$ sufficiently long, on the order a centimeter such that $\lambda \gg \sqrt{\epsilon}$, the effect of the air is negligible and the vacuum model is adequate. 

\section{Summary and conclusions} \label{sec7} 
\textcolor{black}{We have presented a straightforward method to obtain the impedance matching between two acoustic fluids, with the primary emphasis, in this paper, on the water-air interface. The matching layer, or transformer, is achieved by using a solid material like aluminum with no active power source, fluid layers, membranes, or any other passive mechanism. Different from other proposed transformers, including solid ones, the approach in this paper is analytically explicit, resulting in closed-form expressions that relate the performance characteristics, such as the transmission frequency, directly to the material properties or dimensions of the flex-layer.} The proposed flex-layer  acts as an impedance transformer if the system parameters are chosen according to explicit criteria.  Thus, for a given transmission frequency $\omega_0 = 2\pi f_0$ the areal density $m_2$ of the plate facing air must satisfy Eq.\ \eqref{44-0}.  This defines the required thickness of the plate. A second relation, Eq.\ \eqref{top}, defines the required rib spacing $d$, with  $d_0$ of Eq.\ \eqref{059} an approximate over-estimate.  The analytic nature of the acoustic scattering solution along with  asymptotic approximations based on $\epsilon = \frac{Z_a}{Z_w} \ll 1$ leads to explicit expressions such as   Eq.\ \eqref{44-0} and to physical understanding such as the quite distinct motions of the two plates, described in Section \ref{sec6c}.  It also allows us to compare the flex-layer model with a  simple spring-mass transformer defined by an effective mass $m_2$ and effective stiffness $\kappa$ of Eq.\ \eqref{059-}. 

Comparisons of the analytic solution for total transmission  shows excellent agreement with full wave simulations, including for oblique incidence even though the system parameters are chosen to give full transmission for normal incidence, Fig.\ \ref{theta0_E2}.   The effect of air between the plates it to increase the effective stiffness and increase the transmission frequency from that for a vacuum, with the simple approximation of Eq.\ \eqref{888} in good agreement with full wave simulations.  The bandwidth of the transmission resonance depends upon the free system parameters, such as the thickness of the plate facing water.  A parametric study indicates that the Q-factor has a lower achievable limit of $\frac 1{2\sqrt{\epsilon}} = 30.6$, the same as the Q-factor for the ideal spring-mass model of Section \ref{sec2}.   A reduction of  the Q-factor, and associated larger bandwidth, is the subject of a separate study on an alternative transformer model. 

\appendix        

\section{Plate equations from Hamilton's principle} \label{appa} 

The plate equations \eqref{plate1_dis} and  \eqref{plate2_dis} are derived here using Hamilton's principle 
\beq{-12}
\delta \int L \, \dd t= 
\delta \int (T-U+W) \, \dd t= 0
\eeq
with Lagrangian elements   defined by the real-valued displacements and pressure according to  
\beq{34=}
\begin{aligned}
    T &= T_- +T_+ +T_\text{rib},
    \\
    U &= U_- +U_+ +U_\text{rib},
    \\
    W &= W_- +W_+ , 
\end{aligned}
\qquad \text{where} \qquad 
\begin{aligned}
    T_\pm &= \int \frac 12 \rho_\pm h_\pm\, \omega^2 \,  w_\pm^2\, \dd y, 
    \\ 
   U_\pm &= \int \frac 12 D_\pm {w^2_{\pm,yy} }\, \dd y, 
    \\
    W_\pm  &= \mp \int p_\pm w_\pm\, \dd y .
\end{aligned}
\eeq
Here $\pm$ indicates the contributions from the plates on $x=\pm 0$. 
The integrals are over a single period in the $y-$direction that includes one rib between the plates at $y=0$.
This formulation considers the rib as an internal member, and all external forces are contained in the $W_\pm$ terms.

The terms $T_\text{rib}$ and $U_\text{rib}$ are defined by the rib model, and they depend on the plate displacements at $y=0$, that is $w_\pm (0)$.  Taking the variation of  \eqref{-12} with respect to $w_\pm$ yields 
\beq{73=}
\rho_\pm  h_\pm  \omega^2 \, w_\pm  -  D_\pm {w_{\pm,yyyy} }
\mp  p_\pm
 + \frac{\partial  \big(T_\text{rib}- U_\text{rib} \big) }{\partial w_\pm (0)}   \, \delta(y-0)  
 =0   .
\eeq
We consider two rib models that allow us to express $(T_\text{rib}- U_\text{rib}) $ in terms of $w_+$ and $w_-$.

\subsection{A mass-spring rib model }
The rib is a mass $m$ with springs of stiffness $2\kappa$ on either side that attach to the plates, so the static effective stiffness is $\kappa $.  This introduces the mass degree of freedom, $u$, its displacement in the $x-$direction, and 
\beq{36=}
    T_\text{rib} = \frac 12 m \omega^2 { u}^2,
    \qquad
    U_\text{rib} =  \frac 12 2 \kappa \big[ 
    \big(  u - w_-(0)\big)^2 + \big(  u - w_+(0)\big)^2
    \big] .
\eeq
Variation of \eqref{-12} with respect to the rib mass displacement  $u$ leads to 
\beq{37=}
m  \omega^2 u  +2\kappa \big(  w_+(0)  +w_-(0) -2u \big) = 0.
\eeq
Using this to eliminate $u$ gives 
\beq{3-7}
T_\text{rib}  - U_\text{rib}=
\frac {\kappa} 2  \Big[ 
\frac{m  \omega^2}{4 \kappa - m  \omega^2} \big( w_+(0) +  w_-(0) \big)^2
-  \big( w_+(0) -  w_-(0) \big)^2
\Big] .
\eeq
The plate equations \eqref{4=78} then follow from  \eqref{73=}
with
\beq{22-}
Z_{0-} = \frac{\kappa} {-\ii \omega}, 
\qquad
Z_{0+} = \frac{-\ii \omega \kappa m}{4 \kappa - m \omega^2}.  
\eeq

\subsection{The continuous rib model}
The rib is a plate in tension/compression with parameters $\rho$ and $E$ located between $x=-L/2$ and $x=L/2$.   The time harmonic displacement is 
\beq{5+3}
u(x) = \big( w_+(0) + w_-(0) \big)  \frac {\cos kx}{2\cos k\frac L2}
+  \big( w_+(0) - w_-(0) \big) \frac {\sin kx}{2\sin k\frac L2} 
\eeq
where $k = \omega /c$, $c= \sqrt{E/\rho}$, 
and 
\beq{313}
    T_\text{rib} = \int_{-\frac L2}^{\frac L2} \frac 12 \rho h\, \omega^2 \,u^2\, \dd x, 
    \qquad 
   U_\text{rib} =  \int_{-\frac L2}^{\frac L2} \frac 12 Eh\, {u^2_{,x} }\, \dd x        .
\eeq
Hence, 
\beq{377}
T_\text{rib}  - U_\text{rib}= \frac{1} 4 khE\, \Big[ 
\big( w_+(0) +  w_-(0) \big)^2 \, \tan k \frac L2 -  \big( w_+(0) -  w_-(0) \big)^2 \, \cot k \frac L2
\Big] ,
\eeq
and the plate equations  \eqref{4=78} follows from  \eqref{73=}
with  impedances 
\beq{23-}
Z_{0-} =   \ii \rho c \frac h2 \cot \frac{kL}2 , 
\qquad
Z_{0+} =  -\ii \rho c \frac h2 \tan \frac{kL}2 .
\eeq
These are consistent with the spring-mass model for  $\kappa = \frac{Eh}L$ and $m =\rho hL$, as expected for the low frequency regime.   Note  that $Z_{0-}Z_{0+} = \big( \rho c \frac h2 \big)^2 $ which is independent of frequency.

 \section{Alternative derivation of the transmission frequency} \label{appb} 

  We provide a simple and alternative  route to the 
  relation \eqref{429} for the frequency at full transmission.  The derivation does not use infinite sums or explicit solutions in the acoustic media, but relies on the plate equations only.  
  
  Consider a unit period of the layer, Fig.\ (\ref{Flex_thin_thick}).
  Assume the  pressure $p_2$  in the fluid above the layer, air,  acts as a plane wave with particle averaged velocity $\bar v_2$ where $p_2=Z_a \bar v_2$ .    At total transmission the pressure $p_1$  in the water  below the layer is also a wave in one direction  because of zero reflection, and accordingly 
 $p_1=Z_w \bar v_1$.
 We also include the air between the plates which acts as a spring of stiffness $\kappa_a \approx \frac{\rho_ac_a^2}{L_r}$ due to the compressibility of the air, where  $L_r$ is  the plate spacing  \cite{BakhtiaryYekta2024} . 
 The plate equations are then 
  \beq{2+3}
      D_jw_j''''(y) - m_j\omega^2 w_j = (-1)^j \, \big[ - Z_j  \bar v_j
     +\kappa_a (\bar w_1 - \bar w_2) \big] , \quad j=1,2,
 \eeq
 where $Z_1= Z_w$, $Z_2 = Z_a$. 
 Using \ann{$v = -\ii \omega w$} yields 
  \beq{2+13}
    w_j (y)=   w_j^{(0)} u_j(y)  + \frac{ (-1)^j }{ \omega^2 m_j} 
    \big[     -\ii \omega Z_j \bar w_j  +\kappa_a (\bar w_2 - \bar w_1) ]  , \quad j=1,2,
 \eeq
  where $u_j(y)$ are solutions to the homogeneous equations \eqref{2+3} normalized such that 
 $u_j(\frac d2 ) = 1$.   Note, we do not know the precise form of these solutions since the force acting at the rib at $y=\frac d2$ is unknown, except for the fact the forces are equal and opposite on the two ribs. However, we can still find a useful result without knowing $u_1$ and $u_2$. 

 The solutions \eqref{2+13} satisfy two conditions.  The first is the kinematic constraint 
 $w_1(\frac d2 ) =w_2 ( \frac d2 ) $
 implying 
 \beq{5++}
  w_1^{(0)}  + \frac{\ii \omega Z_w \bar w_1  +\kappa_a (\bar w_1 - \bar w_2) }{ \omega^2 m_1}= 
    w_2^{(0)}  + \frac{-\ii \omega Z_a \bar w_2  +\kappa_a (\bar w_2 - \bar w_1)}{ \omega^2 m_2}.
    \eeq
      The displacements $w_j^{(0)}$ can be related to $\bar w_j$ by taking the average 
  of \eqref{2+13},  so that \eqref{5++} becomes  
    \beq{9+9}
\bigg[ \frac{ \ii \omega Z_w }{m_1}  \Big( 1 - \frac 1{\bar u_1}  \Big) 
+  \frac {\omega^2}{\bar u_1} 
 + \frac{ \kappa_a\gamma}{m_1m_2}  
\bigg] \,  \bar w_1 + 
\bigg[ \frac{ \ii \omega Z_a }{m_2}  \Big( 1 - \frac 1{\bar u_2}  \Big) 
- \frac {\omega^2}{\bar u_2} 
 - \frac{ \kappa_a\gamma}{m_1m_2}  
\bigg] \,  \bar w_2 = 0 
\eeq
where 
\beq{2323}
\gamma = m_1+m_2 -  \frac{m_1}{\bar u_2}  -  \frac{m_2}{\bar u_1} .
\eeq
    The second condition is that the shear forces acting at the ribs are equal and opposite: $ D_1w_1''' ( \frac d2 ) + D_2w_2''' ( \frac d2 )= 0$.  The latter 
 is equivalent, by integration from $0$ to $\frac d2$, to taking the average of the sum of the two equations \eqref{2+3}, i.e.
 \beq{3-45}
     (Z_w + \ii \omega m_1)\, \bar w_1 = 
  (Z_a - \ii \omega m_2)\, \bar w_2 . 
  \eeq

    Equations \eqref{9+9} and \eqref{3-45} are then a pair of linear and homogeneous equations in $\bar w_1$ and $\bar w_2$.  In order that non-trivial solutions are possible the determinant must be zero, i.e.
\bal{55}
&  
\Big( \frac{Z_w}{m_1}- \frac{Z_a}{m_2}  \Big) \omega^2 \gamma  -
(Z_w-Z_a) \Big( \omega^2+\frac{\kappa_a\gamma}{m_1m_2}    \Big)
\notag \\
& \qquad
+\ii  \omega\Big\{ 
 \Big( 
\omega^2 + \frac{Z_aZ_w}{m_1m_2}
\Big) \gamma -  (m_1+m_2)  \Big( \omega^2+\frac{\kappa_a\gamma}{m_1m_2}    \Big)   \Big\}  =0.
\eal
The transmission frequency $\omega_0$ follows from \eqref{55} in the same form as \eqref{429}, independent of the air layer stiffness $\kappa_a$.     Note that $\gamma$ also follows from \eqref{55}.  

In summary, the identity  \eqref{429} has been deduced using a lumped parameter model combined with the plate equations for  one spatial period. 



\end{document}